\documentclass[useAMS,usenatbib]{mn2e}
\usepackage{graphicx}
\usepackage{mathptmx}

\def \msol  {\hbox{M$_{\odot}$}}

\def \arcdeg {\hbox{$^{\circ}$}}                        % d
\def \arcmin {\hbox{$^\prime$}}                         % '
\def \arcsec {\hbox{$^{\prime\prime}$}}                 % ``
\def \kms    {\hbox{${\rm km\,s}^{-1}$}}                % km/s
\title[HARP/ACSIS on the JCMT]{HARP/ACSIS: A submillimetre spectral imaging system on the James Clerk Maxwell Telescope}
\author[J.V. Buckle, R. E. Hills, H. Smith et al.]{J.V. Buckle$^{1}$\thanks{E-mail: j.buckle@mrao.cam.ac.uk},
R. E. Hills$^{1,2}$, 
H. Smith$^{1}$, 
W. R. F. Dent$^{3,2}$, 
G. Bell$^{1}$, 
E. I. Curtis$^{1}$, \newauthor 
R. Dace$^{1}$,  
H. Gibson$^{1}$, 
S. F. Graves$^{1}$,  
J. Leech$^{1,4}$\thanks{Present address: Oxford Astrophysics Group, Denys Wilkinson Building, Keble Road, Oxford, OX1 3RH}, 
J. S. Richer$^{1}$, 
R. Williamson$^{1}$\thanks{Present address: Department of Astronomy, Columbia University,  Pupin Physics Laboratories,  550 West 120th Street,  New York, New York 10027, USA}, \newauthor 
S. Withington$^{1}$, 
G. Yassin$^{1}\dagger$, 
R. Bennett$^{3}$, 
P. Hastings$^{3}$, 
I. Laidlaw$^{3}$, 
J. F. Lightfoot$^{3}$, \newauthor 
T. Burgess$^{5}$, 
P. E. Dewdney$^{5}$,  
G. Hovey$^{5}$,  
A. G. Willis$^{5}$, 
R. Redman$^{6}$, 
B. Wooff$^{6}$, \newauthor 
D.S. Berry$^{4}$, 
B. Cavanagh$^{4}$,
G.R. Davis$^{4}$,
J. Dempsey$^{4}$, 
P. Friberg$^{4}$,  
T. Jenness$^{4}$,  \newauthor 
R. Kackley$^{4}$, 
N. P. Rees$^{4}$\thanks{Present address: Diamond Light Source Ltd, Harwell Science and Innovation Campus, Oxfordshire, OX11 0DE},
R. Tilanus$^{4}$,
C. Walther$^{4}$, 
W. Zwart$^{4}$, 
T. M. Klapwijk$^{7}$, \newauthor 
M. Kroug$^{7}$, 
T. Zijlstra$^{7}$\\
$^{1}$Cavendish Astrophysics Group, Cavendish Laboratory, University of Cambridge, J J Thomson Ave., Cambridge CB3 0HE, UK\\
$^{2}$ALMA JAO, Av. El Golf 40 - Piso 18, Las Condes, Santiago, Chile\\
$^{3}$UK Astronomy Technology Centre, Blackford Hill, Edinburgh, EH9 3HJ, UK\\
$^{4}$Joint Astronomy Centre, 660 N. A'ohoku Place, Hilo, HI, 96720, USA\\
$^{5}$Dominion Radio Astrophysical Observatory, PO Box 248, White Lake\\
$^{6}$Herzberg Institute of Astrophysics, 5071 West Saanich Road, Victoria, BC, V9E2E7, Canada\\
$^{7}$Delft University of Technology, Faculty of Applied Sciences, Kavli Institute of Nanoscience, Lorentzweg 1, 2628 CJ Delft, The Netherlands}

\begin{document}
\date{}
\pagerange{\pageref{firstpage}--\pageref{lastpage}} \pubyear{2009}
\maketitle
\label{firstpage}
\begin{abstract}
This paper describes a new Heterodyne Array Receiver Programme (HARP) and
Auto-Correlation Spectral Imaging System (ACSIS) that have recently
been installed and commissioned on the James Clerk Maxwell Telescope
(JCMT). The 16-element focal-plane array receiver, operating in the
submillimetre from 325 to 375~GHz, offers high
(three-dimensional) mapping speeds, along with significant
improvements over single-detector counterparts in calibration and image quality.  
Receiver temperatures are $\sim$120~K across the whole band and system temperatures of $\sim$300K are reached routinely under good weather conditions. The system includes a single-sideband filter so these are SSB figures.
Used in conjunction with ACSIS, the
system can produce large-scale maps rapidly, in one or more frequency settings,
at high spatial and spectral resolution. Fully-sampled maps of size 1
square degree can be observed in under 1 hour.

The scientific need for array receivers arises from the requirement
for programmes to study samples of objects of statistically
significant size, in large-scale unbiased surveys of galactic and
extra-galactic regions. Along with morphological information, the new
spectral imaging system can be used to study the physical and chemical
properties of regions of interest. Its three-dimensional imaging
capabilities are critical for research into turbulence and dynamics. In
addition, HARP/ACSIS will provide highly complementary science
programmes to wide-field continuum studies, and produce the essential
preparatory work for submillimetre interferometers such as the SMA
and ALMA.
\end{abstract}


\section{Introduction}

The James Clerk Maxwell Telescope (JCMT), situated on a high, dry site
near the summit of Mauna Kea, and with a 15~m dish, is the largest
submillimetre observatory in the world. The submillimetre band is rich in
molecular lines, and so high-resolution spectroscopy at these
wavelengths enables studies of space densities, velocities, chemical
structure and excitation in the gaseous material of both galactic and
extra-galactic sources. The majority of these sources are extended on
scales larger than the JCMT beam size of $\sim$14 arcsec (at
345~GHz), and many astronomical targets are much larger, extended on
parsec scales, which frequently need to be mapped to carry out
relevant astronomical research. With submillimetre heterodyne
receivers approaching background-limited performance, the scientific
need for focal-plane heterodyne array receivers is clear. By building
multiple detectors, mapping speeds can be increased, making possible
programmes which observe samples of objects of useful statistical
size.

HARP (Heterodyne Array Receiver Programme) and ACSIS (Auto-Correlation
Spectral Imaging System) have recently been installed and commissioned
on the JCMT. HARP operates in the submillimetre band spanning 325 to 375~GHz,
a frequency range which contains transitions from nearly all the most abundant molecules
in interstellar gas.  The
key scientific programmes that are expected to be carried out with
the new system are
\begin{itemize}
\item surveys of the distribution and properties of molecular clouds in face-on
and edge-on galactic disks, starburst and Seyfert galaxies, and
interacting galaxy systems;
\item large-scale unbiased surveys of star formation and outflows in
molecular clouds;
\item large-scale studies of hierarchical structure and clumping in
molecular clouds;
\item studies of the chemical processing of molecular clouds by shocks
and turbulence;
\item surveys of the dynamics and excitation of molecular gas in the Galactic centre;
\item deep, narrow surveys of molecular gas in Solar System objects;
\item studies of the chemical and physical state of comets near perihelion.
\end{itemize}

Several large JCMT legacy survey projects have been developed to exploit this new instrumentation in the above scientific areas \citep{ward,plume,matthews,wilson}. The combination of HARP and ACSIS makes possible science
which is highly complementary to the wide-field continuum studies of
cold dust in distant galaxies and the earliest stages of star
formation that have been carried out by SCUBA \citep{hollanda}, and will continue with
SCUBA-2 \citep{hollandb}. It will also provide the essential scientific preparatory
work for submillimetre interferometers such as the SMA and ALMA.

In addition to the increase in mapping speeds, HARP/ACSIS offers
supplemental benefits, the most important of which is probably the
decrease in calibration errors and increase in image quality, since
 multiple detectors observe the source simultaneously. Relative
calibration of individual detectors is more accurate, as data are
taken simultaneously through the same atmospheric path. By making
overlapping maps, pointing drifts between maps can be removed. For
objects smaller than the array field of view, edge detectors can be
used to make reference spectra through on-array beam switching by using
appropriately selected chop and node distances, thereby increasing the
mapping speed of compact objects.

The full ACSIS capabilities are described in Sec.~\ref{sec:acsis}. When used with HARP, ACSIS offers either wide bandwidths (up to $\sim$1.9~GHz
for each of the 16 IF channels),
or high spectral resolution (with a channel spacing as
small as 31~kHz, or 0.03~\kms). In addition, it
can provide one or two sub-bands per IF channel,
allowing simultaneous observations of
multiple lines within the HARP 5~GHz IF frequency. One example is
observing the $J$=3$\rightarrow$2 lines of $^{13}$CO and C$^{18}$O with two separate
high-resolution sub-bands.
ACSIS and the upgraded JCMT Observatory Control System allow for rapid
observing and data taking. Fast scanning enables data to be taken
continuously at up to 10~Hz, and it is possible to make
fully-sampled maps with HARP of 1 square degree in less than 1 hour.

\begin{figure}
\centering
\vbox{
\noindent \includegraphics[width=8cm]{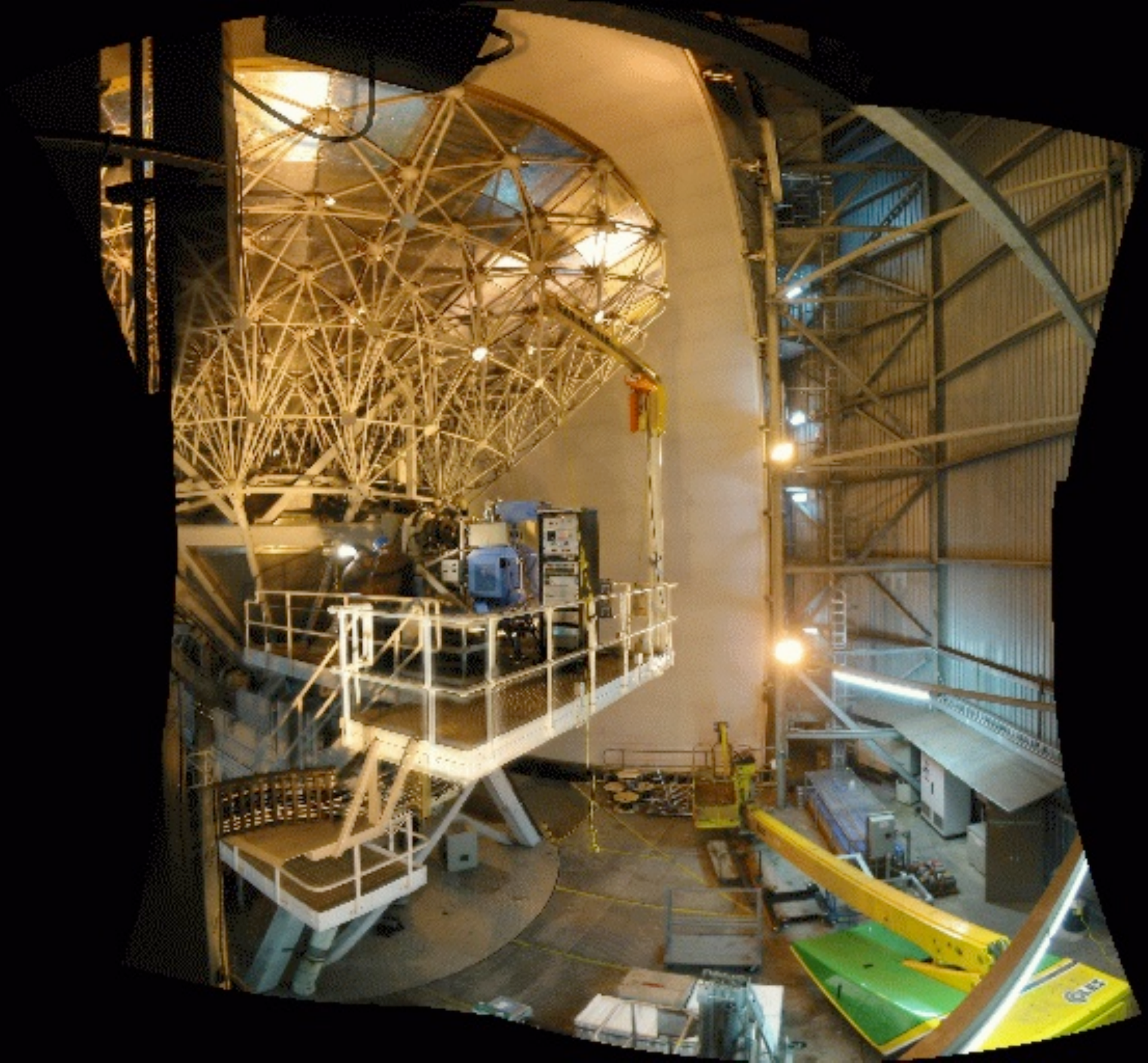}
\noindent \includegraphics[width=8cm]{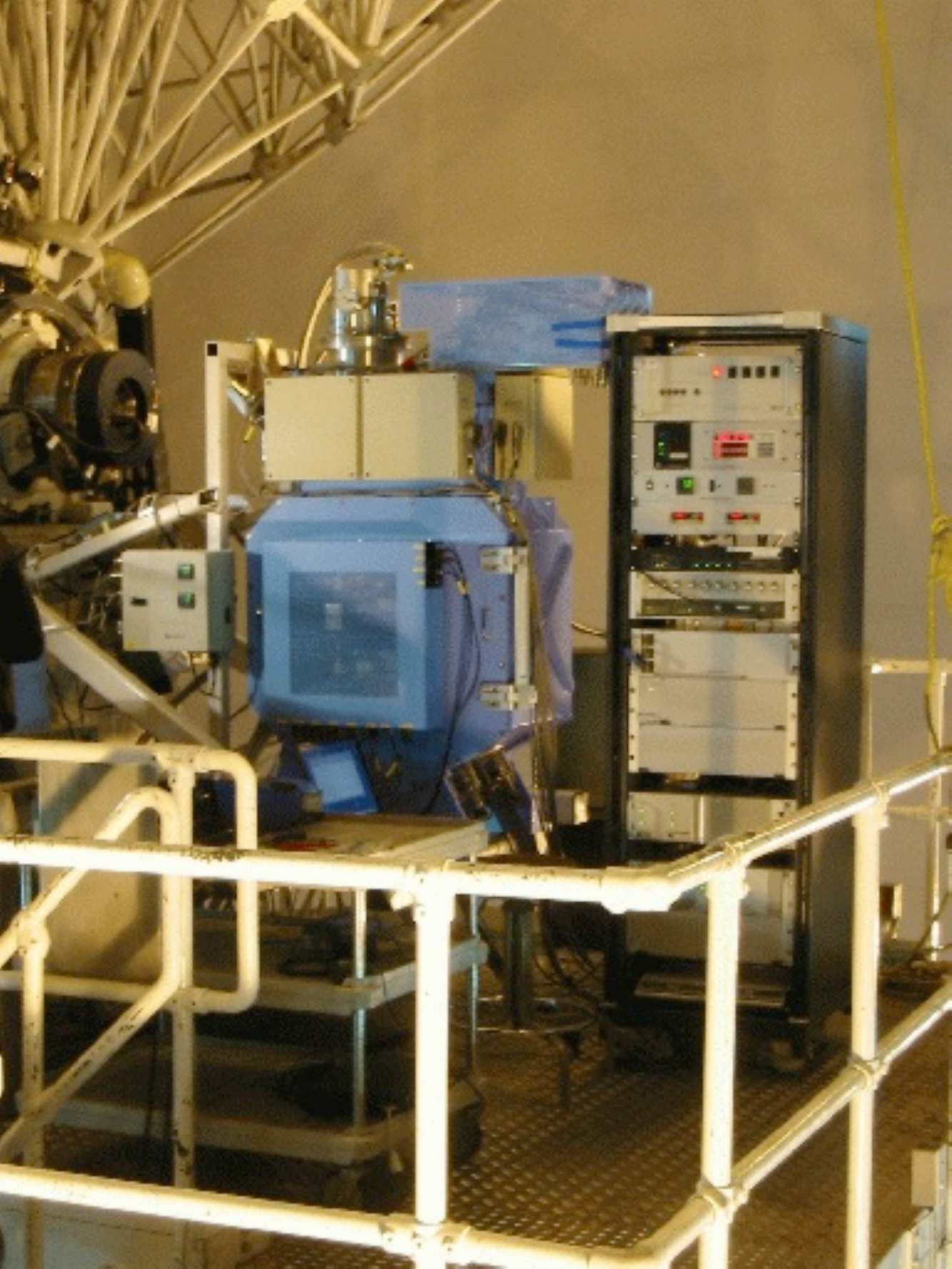}}
\caption{\label{nasmyth} HARP shown on the right-hand Nasmyth platform of the JCMT, and a zoomed-in image of HARP and the instrument electronics box.}
\end{figure}

\begin{figure}
\centering
\noindent \includegraphics[width=6cm,angle=-90]{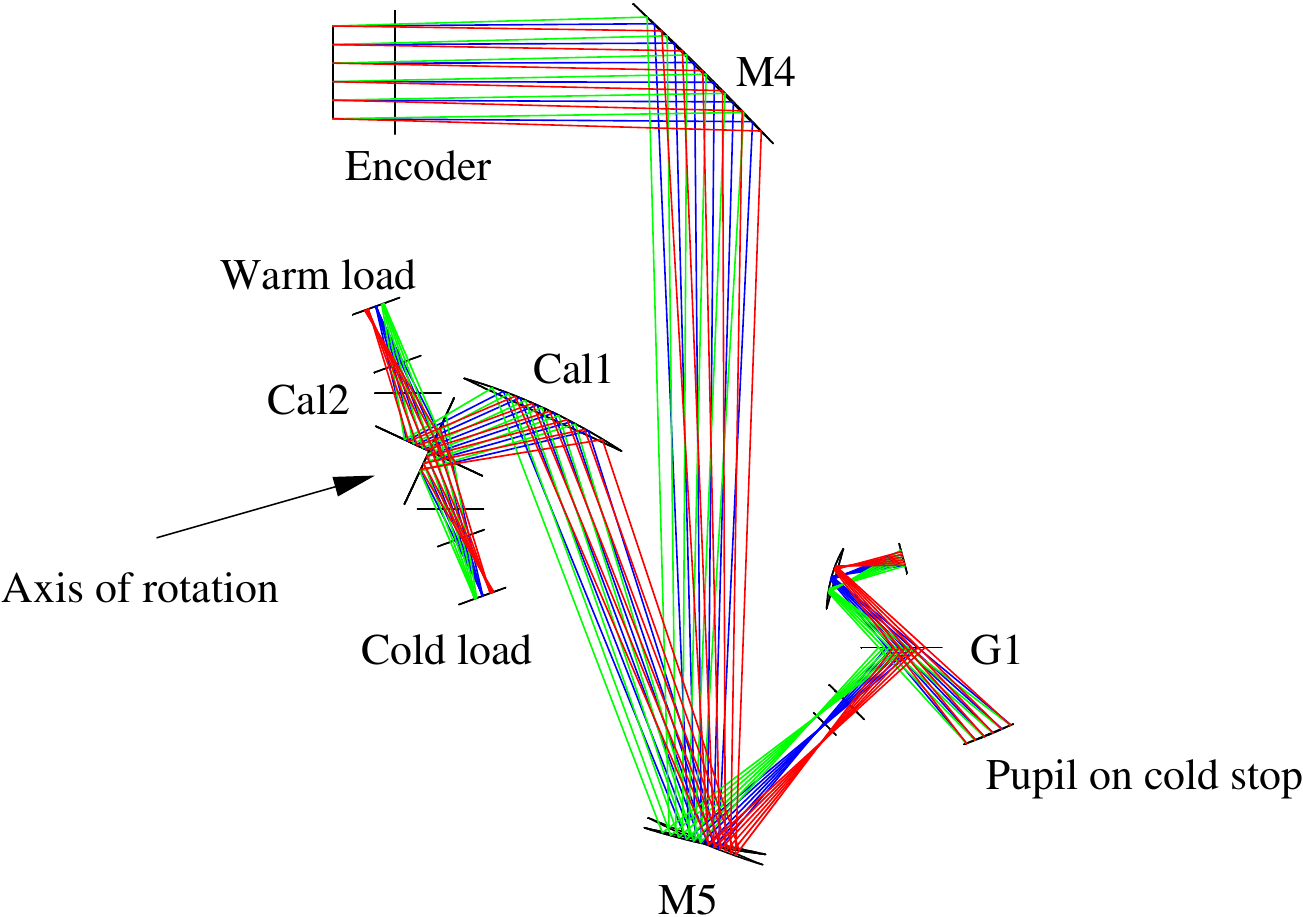}
\caption{\label{relayoptics} Diagram of the relay optics from the encoder to the cold stop, through a series of labelled mirrors.}
\end{figure}

\begin{figure}
\centering
\noindent \includegraphics[width=6cm,angle=-90]{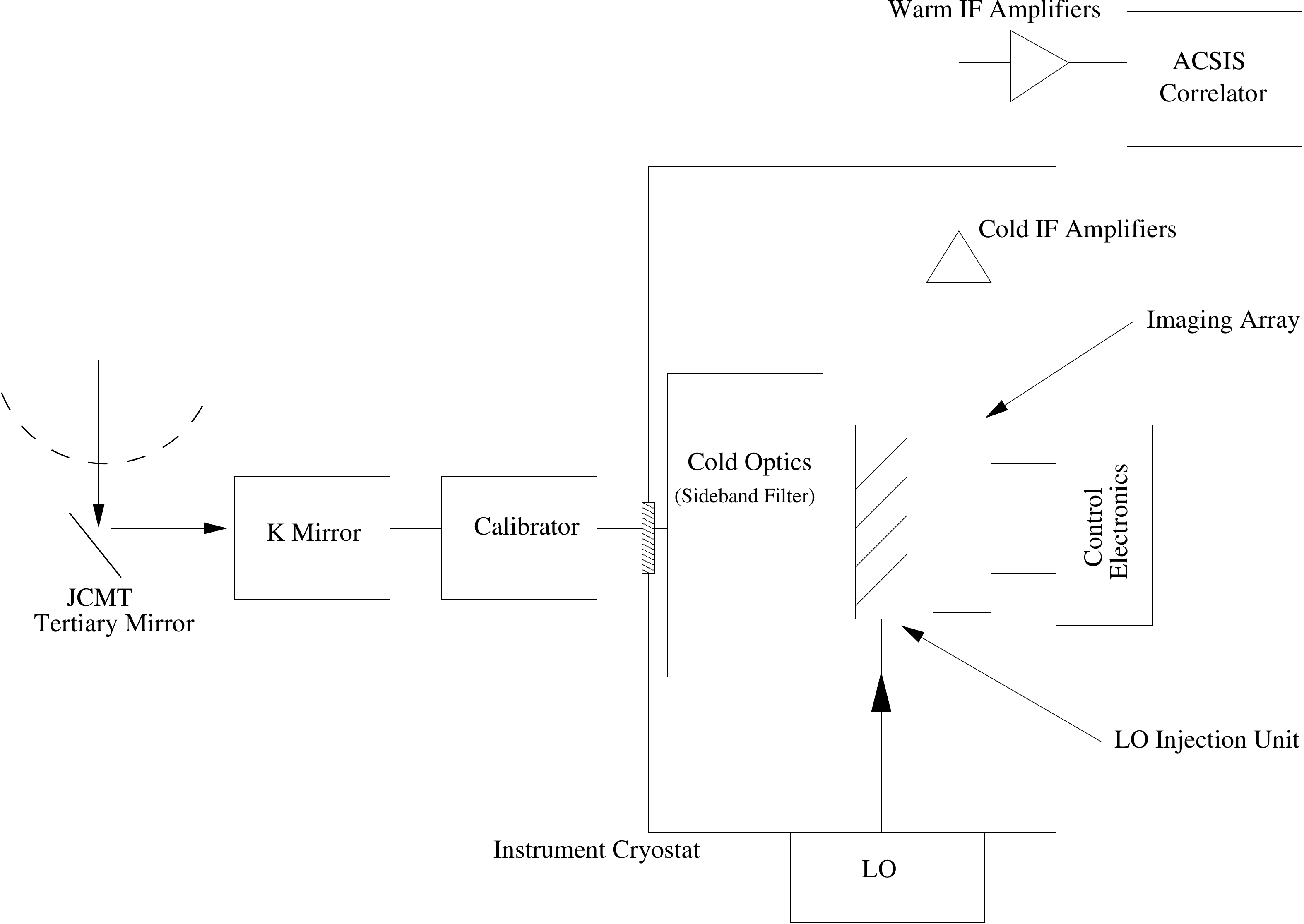}
\caption{\label{harp_schem} A schematic diagram of the components of the HARP receiver system, adapted from \citet{leech}.}
\end{figure}

As shown in Fig.~\ref{nasmyth}, HARP has been installed on the
right-hand Nasmyth platform of the JCMT. An optical relay
(Fig.~\ref{relayoptics}) provides the field of view and imaging
performance to match the HARP specifications. The relay optics also
bring the beam down to a lower level so that the cryostat is
conveniently accessible on the Nasmyth platform. A schematic diagram of the
components of HARP is shown in Fig.~\ref{harp_schem}. The path is further
explained in Sec.~\ref{sec:optics}.  Each of the
components is described in the following sections, with Sec.~\ref{sec:harp} describing the HARP components, and Sec.~\ref{sec:acsis} describing the ACSIS components. 

Designing and building HARP was a collaborative project between the
Cavendish Astrophysics Group in Cambridge, the UK Astronomy Technology
Centre in Edinburgh, the Herzberg Institute for Astronomy in Victoria,
Canada, and the Joint Astronomy Centre in Hawaii, with  the Kavli Institute of Nanoscience at Delft contributing the critical SIS junctions.  ACSIS was built in collaboration between the Dominion
Radio Astrophysical Observatory in Penticton, Canada, the UK Astronomy
Technology Centre and the Joint Astronomy Centre.

\section{HARP design concepts}
\label{sec:harp}

A description of the key design features of the instrument, along with
technical and organisational overviews, have previously been
published \citep{smitha,smithb}. In this section we present a brief
overview of the key features of HARP's specification, design and performance.

\subsection{Overview}

\begin{figure}
\centering
\noindent \includegraphics[width=8cm,angle=-90]{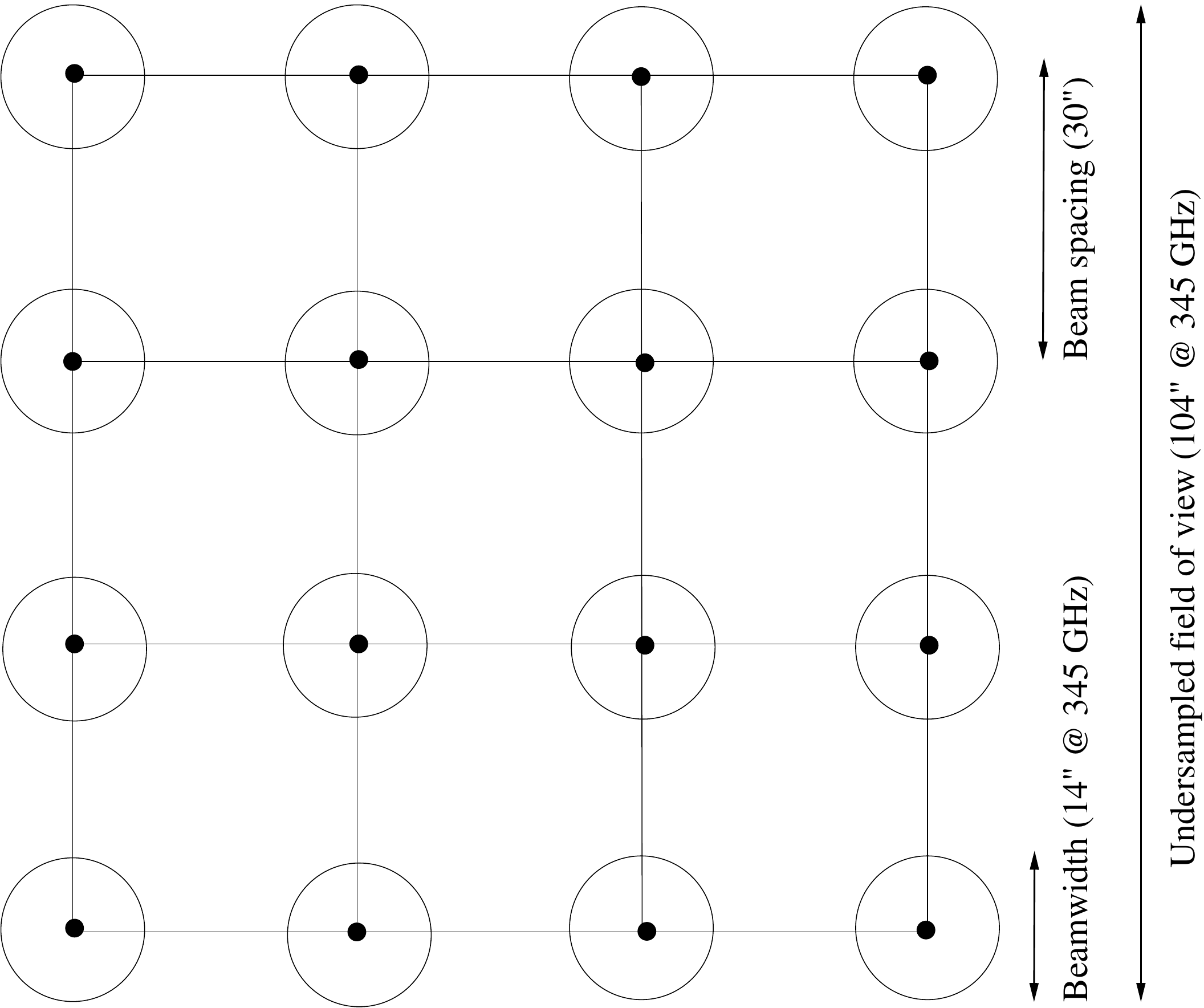}
\caption{\label{harplayout} Schematic diagram of the HARP grid layout.}
\end{figure}

HARP comprises 16 detectors laid out on a 4$\times$4 grid, with an on-sky projected beam separation of 30~arcsec. 
At 345~GHz the beam size is 14~arcsec, and the under-sampled field
of view of HARP is 104~$\times$~104~arcsec, as shown in
Fig.~\ref{harplayout}.  In a single pointed observation, the map is
under-sampled by a factor of 4.4 in area, and by a factor of 17.6 with
respect to the Nyquist frequency $(\lambda/\rm 2D)$. 

Beam rotation, using a new K-mirror (Sec.~\ref{sec-kmirror}), optimizes
the observing efficiency. For objects similar to or smaller than the array
field of view,  beam rotation ensures no edge detectors miss the source
emission. For larger objects, and particularly for scan mapping, the
array can be orientated with respect to the scan direction to provide
good sampling (Sec.~\ref{sec-obsmodes}). HARP is easily and rapidly tuned across the operating range of
325 to 375~GHz using automated control software in 10--30
seconds (Sec.~\ref{sec-controlsw}). The IF frequency is 5~GHz, and, in conjunction with
ACSIS, HARP offers up to 1.9~GHz of bandwidth.

HARP uses single sideband (SSB) filtering to minimize
the system temperatures and improve calibration accuracy. HARP
sensitivities have exceeded expectation, with the combination of
receiver noise temperature and beam efficiency better than the required 330~K
(SSB).  Relative inter-receptor
calibration, measured on continuum sources, is accurate to better than 5~per cent, and relative beam positions can be
measured to an accuracy of 1~arcsec.

HARP utilizes a number of innovative features across all elements of
the design. The key features of these are outlined below. Further
technical details and schematic diagrams are presented elsewhere \citep{smitha,leech,williamson,smithb,bell}.

\subsection{Optical system}
\label{sec:optics}

The optical system for HARP has three sections. The K-mirror is inside
the receiver cabin and described in Sec.~\ref{sec-kmirror}. The warm
optics, comprised of the relay optics and calibration system, are
exposed on the Nasmyth platform, and are described in
Sec.~\ref{sec-warmoptics}. The cold optics are contained inside the
cryostat, and are described in Sec.~\ref{sec-coldoptics}.

\subsubsection{K-mirror}
\label{sec-kmirror}

\begin{figure}
\centering
\noindent \includegraphics[width=1.5cm,angle=-90]{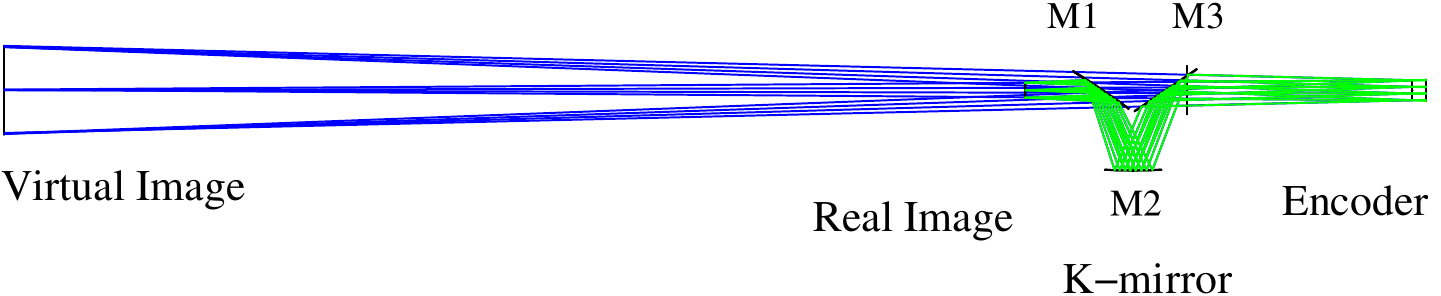}
\caption{\label{kmirror} Schematic of the K-mirror, showing the three mirrors, M1 to M3. 
The JCMT tertiary mirror is just to the left of the real image.}
\end{figure}

\begin{figure}
\centering
\noindent \includegraphics[width=10cm,angle=0]{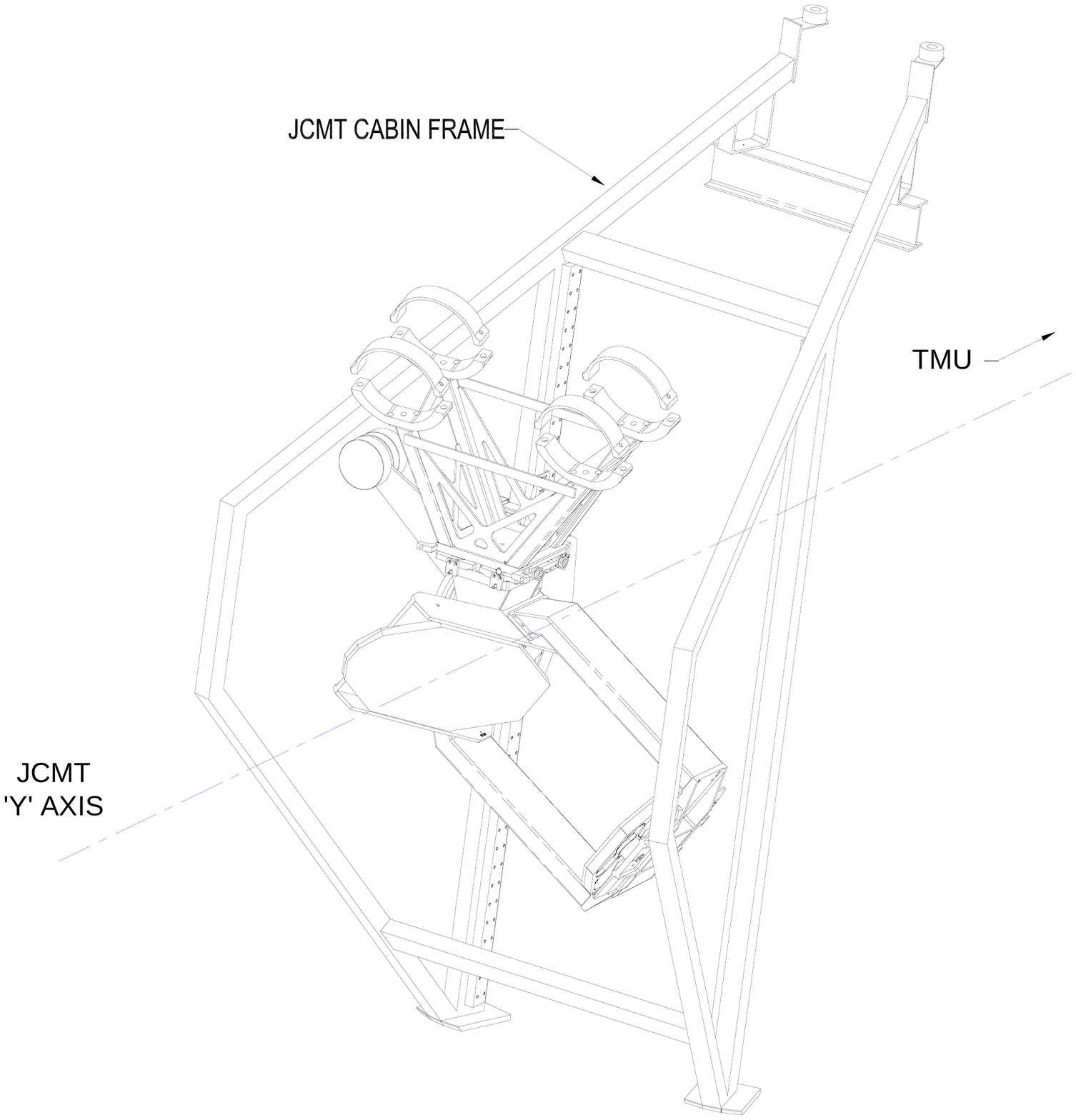}
\caption{\label{kmirror2} Schematic overview of the K-mirror in the JCMT receiver cabin. The JCMT `Y' axis is the elevation axis.}
\end{figure}

The K-mirror (Fig.~\ref{kmirror}) consists of three large powered
mirrors with modest curvature.  The system acts as an image and
polarization rotator, and forms an image of the secondary mirror at a
point near the elevation encoder.  An overview of the K-mirror in the JCMT receiver cabin is shown in Fig.~\ref{kmirror2}. The K-mirror
also maximizes the field of view available to instruments on the
Nasmyth platform, which otherwise would be severely limited due to the
relatively small hole in the encoder axis.  The K-mirror rotates as a
unit about the elevation axis of the telescope, and has a rotational
range of $\pm$57.5 degrees, limited through control software to $\pm$55
degrees, with the effective rotation of the image and polarization
twice the K-mirror rotation angle. The K-mirror can rotate at a rate of 5 degrees~s$^{-1}$,
with a rotational position accuracy of $\pm$0.1 degree. The design
allows for a field of view of $\sim$200~arcsec. The system is
designed to have very low losses and aberrations. Spillover loss
is designed to be less than 1~per cent, and in fact is
expected to be $<$ 0.5~per cent at 850~$\micron$. The losses per mirror
should be less than 0.5~per cent at this wavelength, which will contribute $\sim$3~K
to the system temperature from each mirror. In order to meet the needs of possible future JCMT instrumentation,
the mirror surfaces have been finished so that they are compatible
with the highest operational frequencies at the JCMT of $\sim$870~GHz.

Rotation of the K-mirror is controlled by the Telescope Control System
(TCS). The TCS commands a continuous rotation of the K-mirror to
compensate for the change in parallactic angle of the source and the
rotation of the elevation axis of the telescope. When the telescope is
commanded to slew to a new position, the TCS computes an optimal
orientation of the focal plane relative to the tracking coordinate
system to give the longest tracking time on source before hitting one
of the K-mirror rotation limits. This optimal orientation corresponds
to an initial K-mirror angle to which it is slewed and then the TCS
continuously updates the angle to maintain a fixed orientation between
the focal plane and tracking coordinate systems throughout the
observation.

The pointing offsets due to possible misalignment of the K-mirror axis
and its axis of rotation have been calculated, and included in the
JCMT pointing model.  The effective pointing of the telescope on the
sky is changed if the mirrors are displaced or tilted, so a
requirement of the K-mirror is that total pointing errors are less
than 10~arcsec. Residual pointing errors, after subtraction of the
known correction terms, are a few arcsec, partly due to the number of new terms
in the pointing model introduced by the K-mirror.

\subsubsection{Calibration}
\label{sec-warmoptics}

\begin{figure}
\centering
\noindent \includegraphics[width=7cm,angle=0]{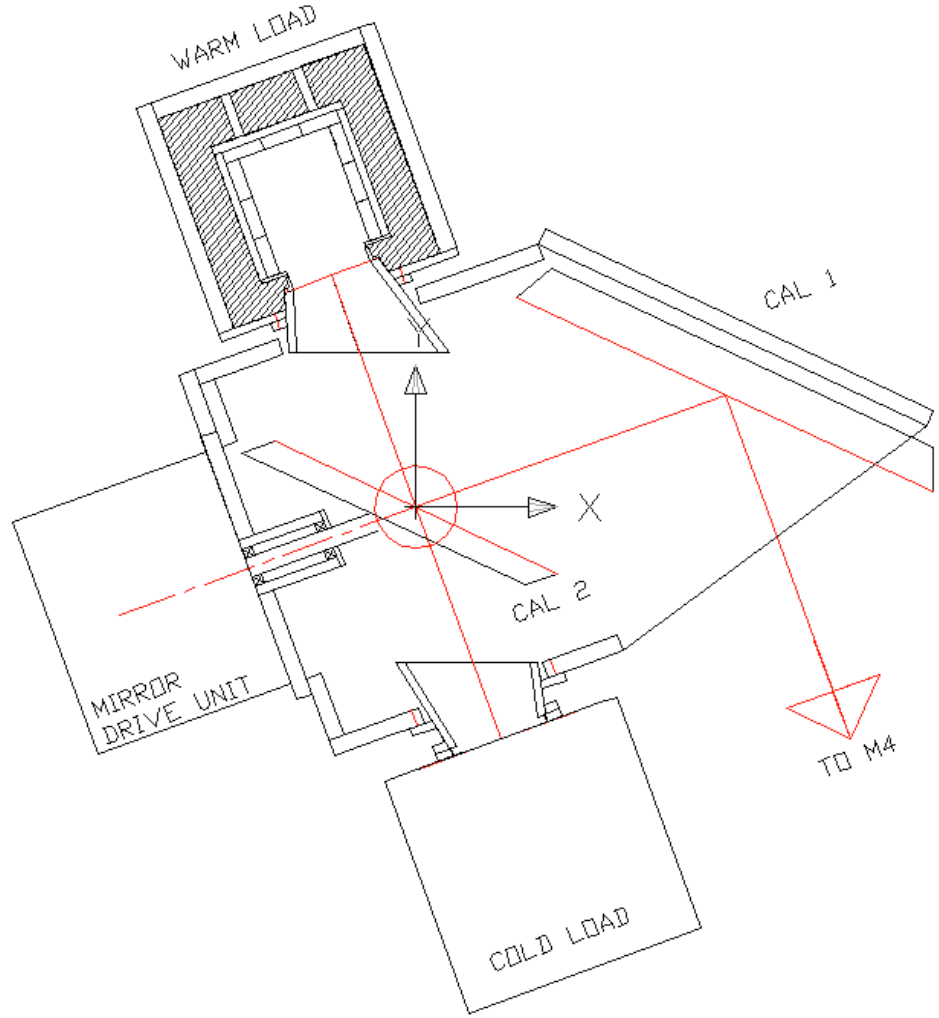}
\caption{\label{calunit} The calibration system \citep[from][]{williamson}. }
\end{figure}

A schematic overview of the calibration system is shown in
Fig.~\ref{calunit}. The calibration unit was built as an independently
mounted unit containing the two mirrors, Cal1 and Cal2, the warm and
cold loads, and a spectral-line calibration signal
(Fig.~\ref{relayoptics}). Cal1 is strongly concave and fixed, while
Cal2 is flat, and can turn from the warm load to the cold load in
under 2 s.  The loads are isothermal cuboid cavities lined with
absorbing tiles that have high efficiency at the HARP operational
frequencies. A thermally isolated conical reflecting baffle at the
input limits the IR coupling. The positions of the
loads were chosen to minimize convection. 

To make calibration measurements, a lower relay mirror (M5,
Fig.~\ref{relayoptics}) is tilted mechanically to point to the
calibrator unit, where the Cal2 mirror focuses the beams on to either
the warm or cold load. The Cal2 mirror is actuated by a standard
four-position ``Geneva'' mechanism, while the M5 mirror is actuated by a
novel Geneva design of a two-position type. The Cal1 mirror then rotates
to switch between the loads. The warm load is heated by resistors to
temperatures $\sim$40~K above ambient. The cold load is cooled to $\sim$10~K below
ambient temperatures using a Peltier cooler, with the intention of
providing a load that is at approximately the same temperature as the water vapour
in the atmosphere. Two- or three-load
calibration measurements can be taken in a few seconds. A two-load
calibration, which measures the power from the sky and the power from
the cold load, should then accurately remove atmospheric
attenuation. The three-load calibration, which additionally measures
the power from the warm load, provides a measure of the receiver
temperature.

The Cal2 mirror has a third position which allows it to point at the
spectral-line calibration signal. This coherent signal is used to
check the tuning of the interferometer (Sec.~\ref{sec-coldoptics}) and to optimize the
signal-to-noise ratio when adjusting tuning parameters such as the
bias voltage and local oscillator (LO) level. It uses a YIG oscillator and a multiplier
to produce a narrow line which can be tuned to lie anywhere in the
HARP operational frequency range. The optics are arranged to couple this signal
into all the mixers at the same time with a reasonable degree of
uniformity.

The night-to-night calibration accuracy has been measured to be within
JCMT guidelines of 20~per cent, and is usually better than this
(Sec.~\ref{sec:calibration}). Work on improving the accuracy and
stability of the combined HARP, ACSIS, telescope and software systems
continues at the JCMT. The receptor-to-receptor calibration accuracy
was measured with observations on the nearly full Moon (Fig.~\ref{moon}), and indicate maximum
differences between levels less than 5~per cent. Any variation of calibration
with ambient temperature is at levels below both the night-to-night
and receptor-to-receptor calibration accuracy.

\subsubsection{Cold optics}
\label{sec-coldoptics}

\begin{figure}
\centering
\noindent \includegraphics[width=9cm,angle=0]{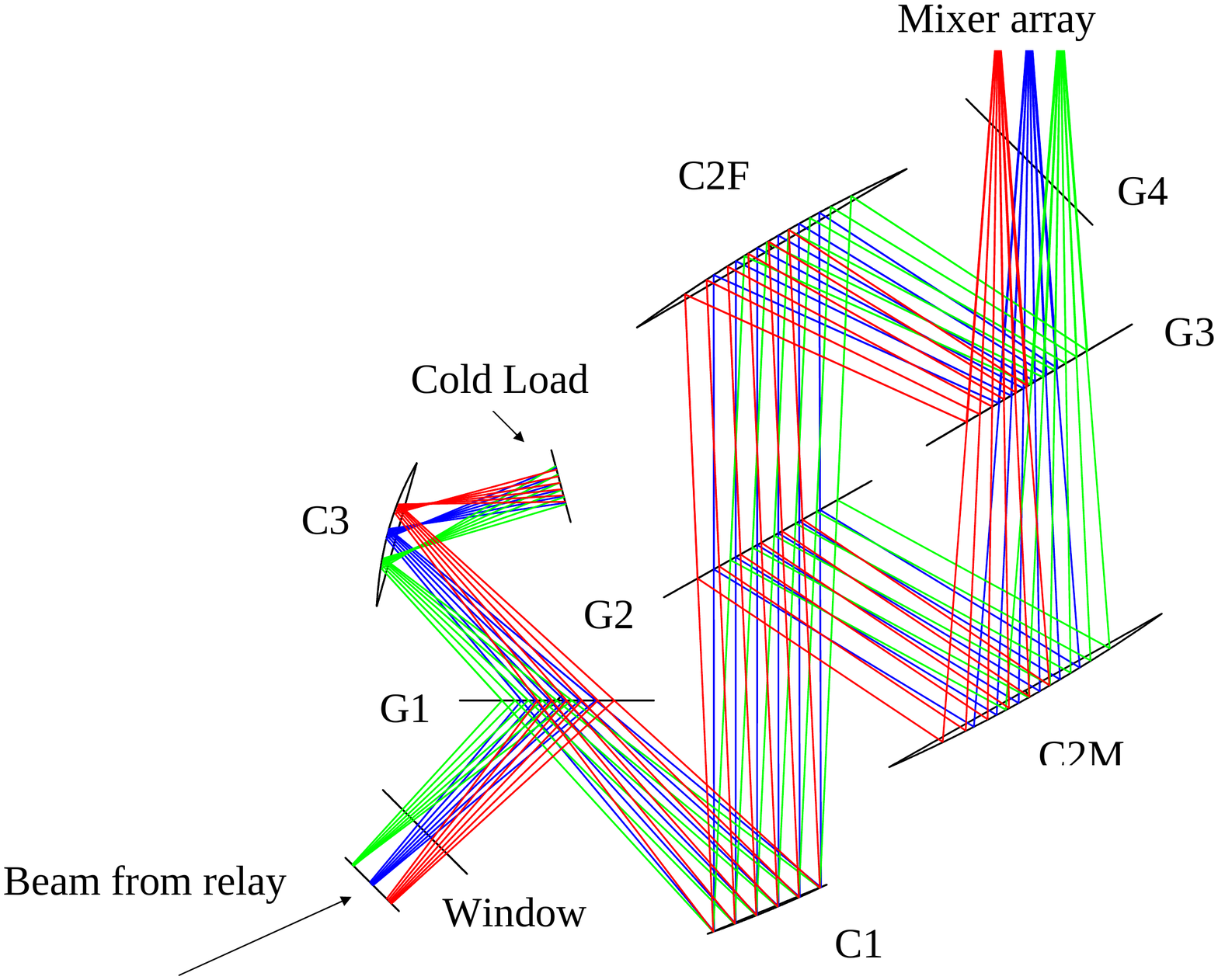}
\caption{\label{coldoptics} Overview of the cold optics, showing the four mirrors (Cn) and the four grids (Gn) of the polarizing Mach-Zehnder interferometer.}
\end{figure}

The cold optics are located inside the cryostat, and operate at
temperatures $\sim$60~K.  Four powered mirrors
(Fig.~\ref{coldoptics}) between the encoder and the array are optimized
together to take advantage of aberration balancing. Utilizing a
reflective, slightly curved cold stop inside the cryostat, the system
forms an image of the sky with low aberration at the detector
array. The smallest mirror, C1, forms the cold stop, while C2F and C2M
create an image of the sky at the mixer array. 

A polarizing Mach-Zehnder interferometer is
used for the sideband separation and the image sideband is terminated in a cold
load, inside the cryostat. C2M is moved to tune
the interferometer, and the full range of movement for this tuning can
be achieved in under 10~s.  The HARP IF frequency of
5~GHz sets the path difference requirement for the two beams through
the interferometer of 15~mm. Leakage and cross talk levels between the
beams were looked for during on-sky commissioning, and were not
detected at the 1~per cent level.

The interferometer is tuned so that the
path difference is a whole number of wavelengths for the image
frequency.  This means that the path difference of the signal
frequency, once it has completed transmission through the system, is
an odd number of half-wavelengths. This results in the signal in the
desired sideband being coupled to the mixers and in the other sideband
the mixers are coupled to the SSB load. The tuning range achieved
covers 324~GHz to 376~GHz. The grids are made from gold-plated
tungsten wire with diameter 10~$\micron$ and spacing 25~$\micron$.  G1 and G4 are rectangular, while G2 and G3
are circular. The two sets of grids are aligned to split the signals
into $\pm$45 degree components.  The image sideband dump provides a
termination for the beam transmitted through G1, and consists of a
concentrator mirror, C3, which operates at $\sim$60~K, and a thermally-isolated cold load (the SSB load), which is cooled to $\sim$12~K. The details of the design and laboratory testing
of the cold optics systems are fully explained in \cite{williamson} and
\cite{bell}.

The reflective cold stop, C1 is surrounded by cold absorber, which is
typically at 18~K. The cold optics also truncate the side-lobes of the
feed pattern, so they see cold absorber inside the cryostat rather
than thermal emission from the region around the elevation encoder.
The interferometer is the critical item in the cold optics, and
provides the datum position to which the other components are
referenced, with the cryostat window providing the interface to the
rest of the system. Sideband separation commissioning tests were
carried out across the observing bandwidth, and were measured to be
better than 19 dB on average (Sec.~\ref{sec-sysobs}).

\subsection{Imaging array}

\begin{figure}
\centering
\noindent \includegraphics[width=9cm,angle=0]{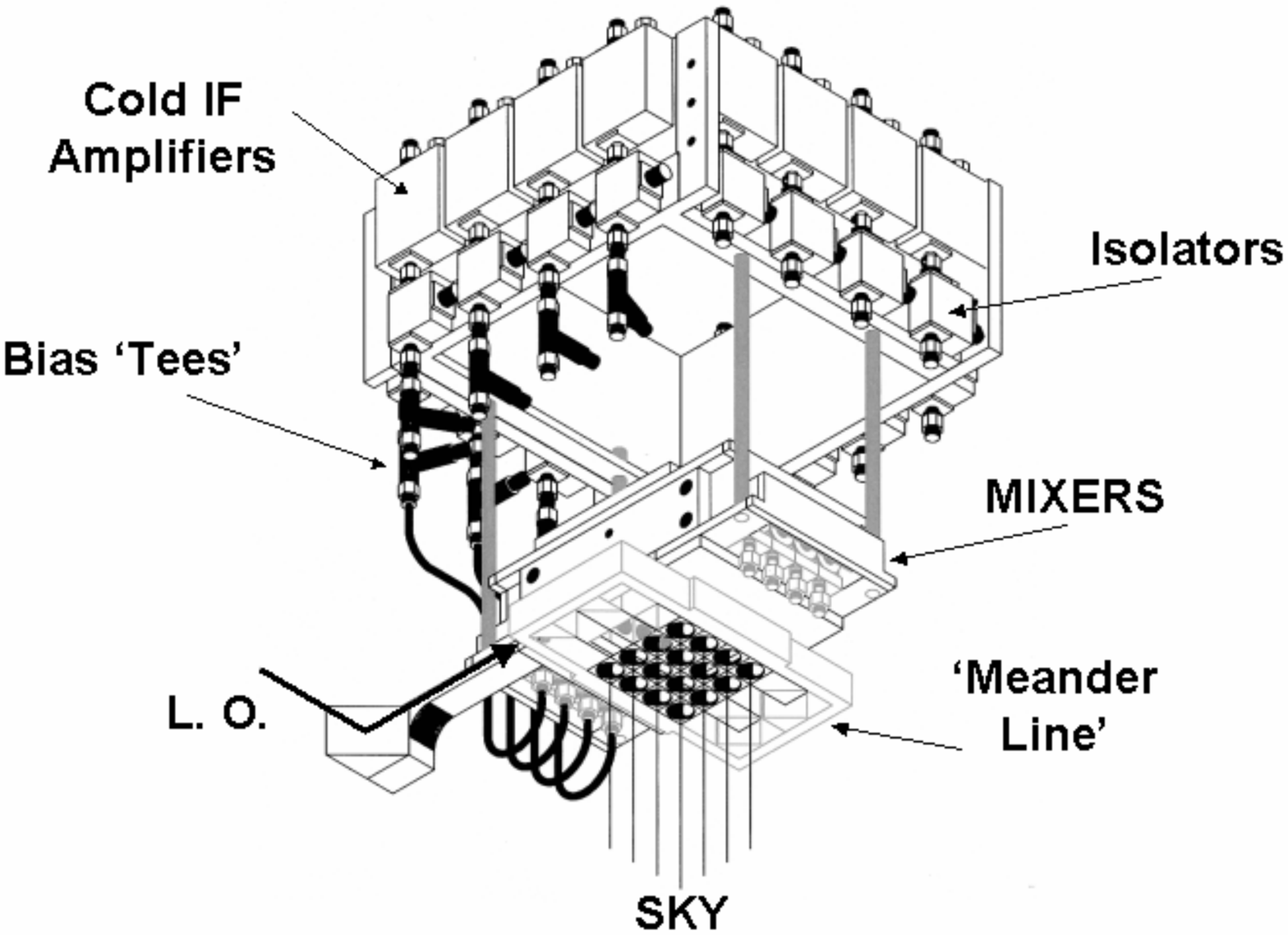}
\caption{\label{array} The HARP imaging array.}
\end{figure}

 The imaging array unit comprises an array of mylar beam splitters for
 LO injection, two decks of eight horn-reflector
 antennas holding 16 SIS mixers with air-cored coils for Josephson
 current suppression, and 16 HEMT cold IF amplifiers with isolators
 and bias tees (a T-shaped multiplexer).  The horn reflector antennas
 consist of split-block, directly machined corrugated horns with
 ellipsoidal reflectors. The corrugated horns were designed to be easy
 to mass produce using direct machining into a split block. The novel
 horn design features constant depth corrugations and delivers good
 beam patterns across the required bandwidth while avoiding difficult
 to machine $\sim \lambda/2$ deep corrugations near the throat of the
 horn. Ellipsoidal reflectors were chosen to provide efficient
 coupling of the horns to both the telescope fore-optics and the LO
 meander line. The imaging array unit is shown schematically in
 Fig.~\ref{array}. The array is mounted inside the 20~K heat shield.

\subsubsection{The local oscillator coupler}

\begin{figure}
\centering
\noindent \includegraphics[width=6cm,angle=-90]{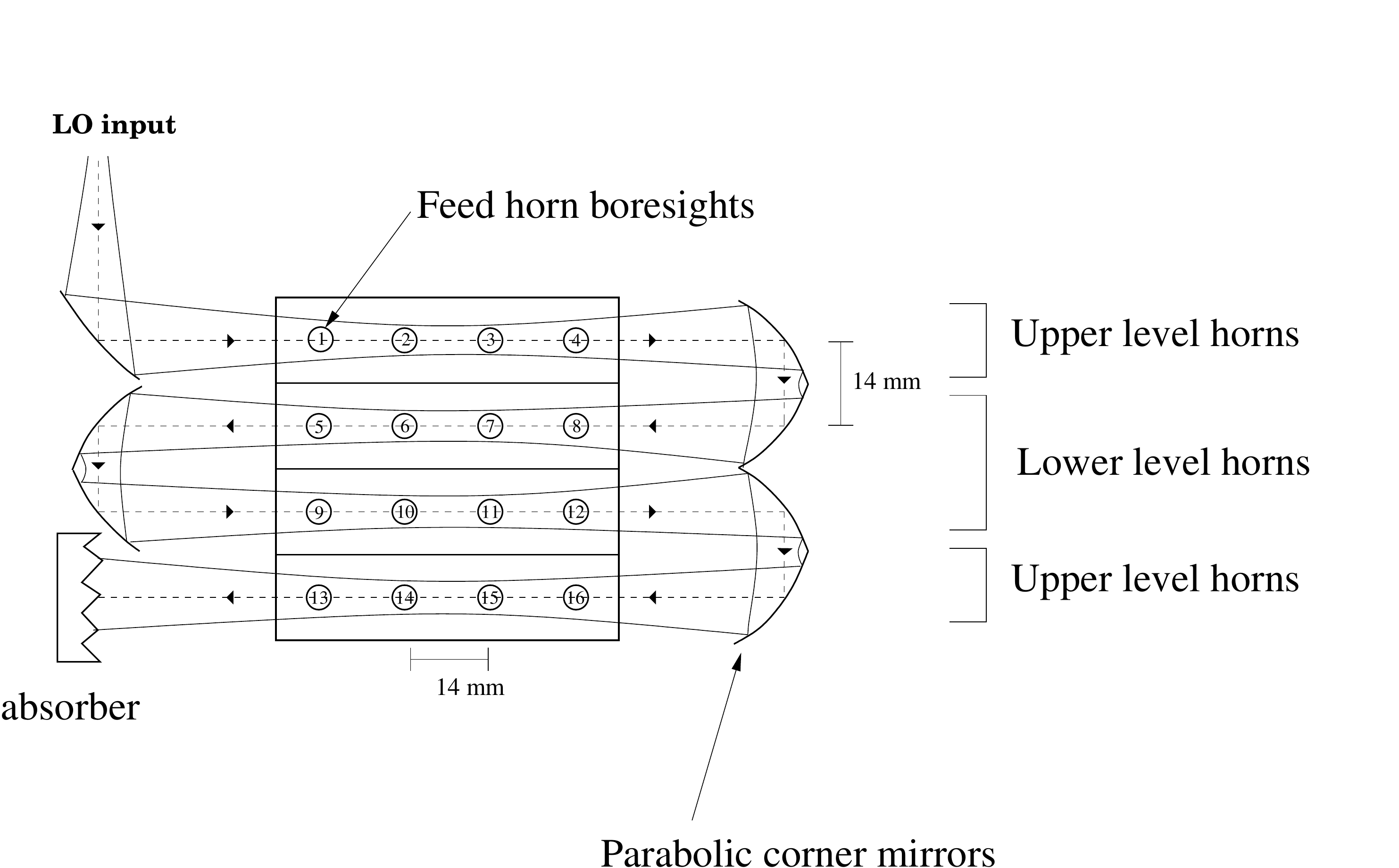}
\caption{\label{meander} Plan view of the LO coupler.}
\end{figure}

 The LO coupler uses a `meander line' to provide the efficient
 coupling necessary to pump 16 mixers (Fig.~\ref{meander}). At each
 crossing point there is a 45 degree beam-splitter made of thin mylar
 film. The mylar in the beam-splitters was pre-tensioned to ensure it
 remains tight when cooled. In order to prevent the beam from
 diverging due to diffraction as it passes through the meander line,
 the roof mirrors that are used to turn the beam around after passing
 in front of each set of 4 mixers consist of off-axis
 paraboloids. These produce beam waists at the mid-points of each
 path. This arrangement has a large RF bandwidth, and injects LO power
 in a highly efficient manner \citep{leech}.

\begin{figure}
\centering
\noindent \includegraphics[width=8cm]{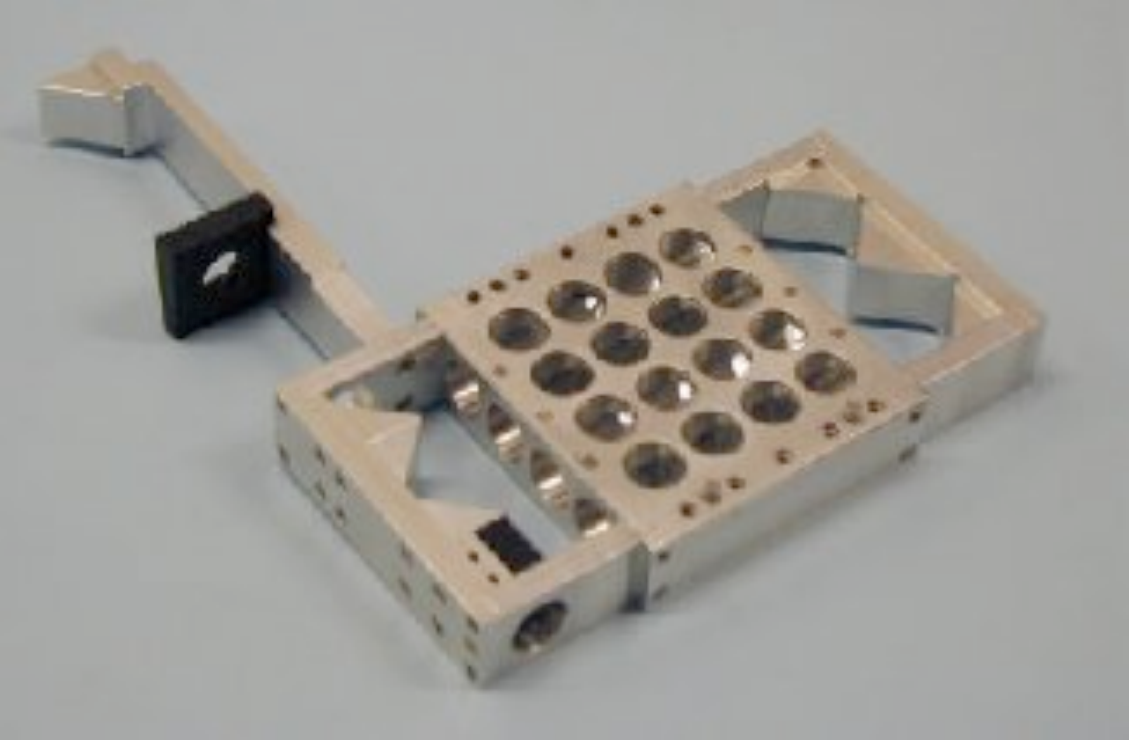}
\caption{\label{locoupler} Photograph of the assembled LO coupler, also showing one
mirror of the LO relay and the stop. }
\end{figure}

\begin{figure*}
\centering
\noindent \includegraphics[width=12cm]{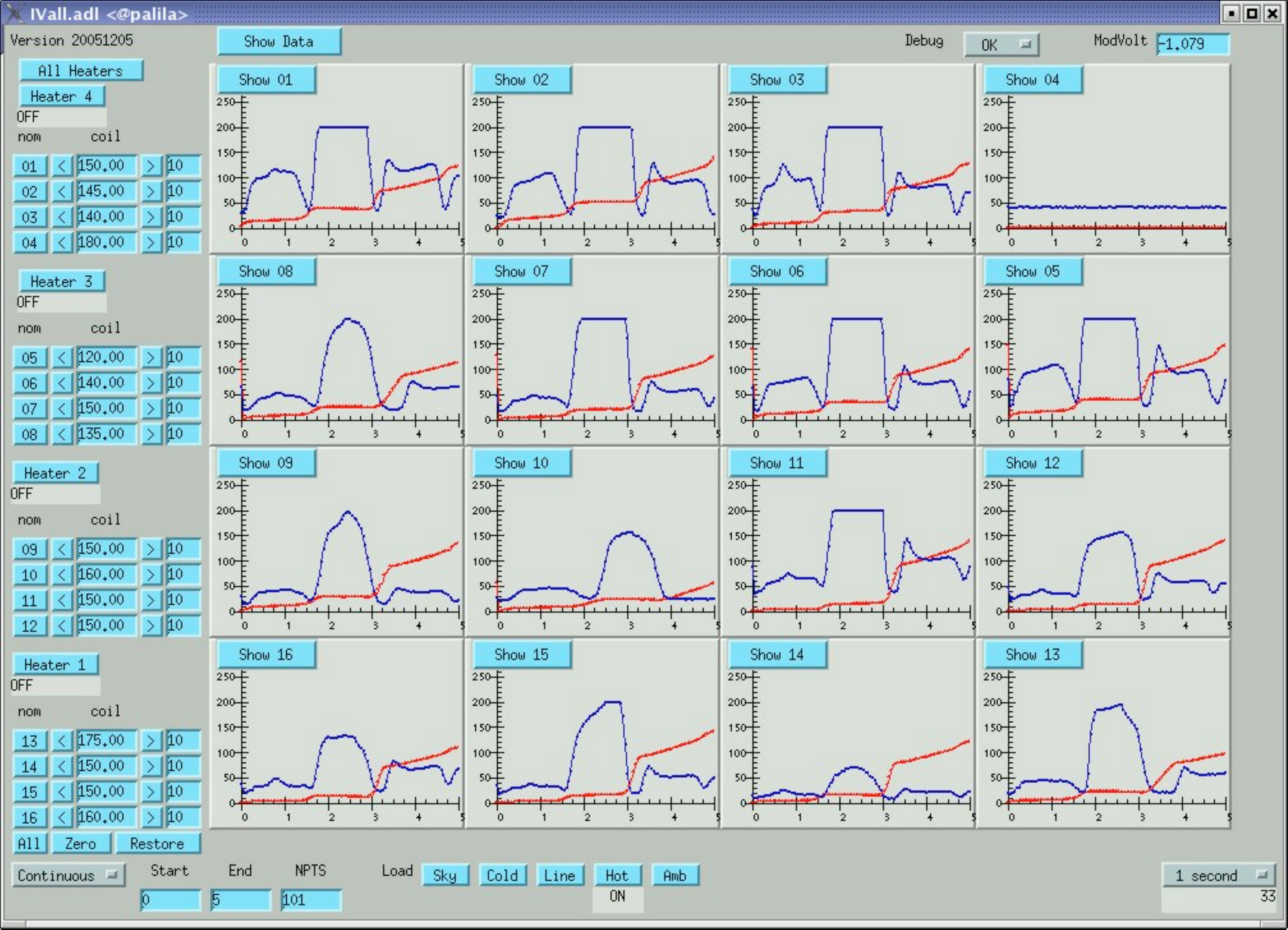}
\caption{\label{ivcurves} The HARP control system I/V and P(IF)/V curves display.}
\end{figure*}

For ease of maintenance and upgrades, this unit is independent of the
cold optics, with the LO plate mounted on the door of the
cryostat. The LO is generated outside the cryostat (Fig.~\ref{harp_schem}), and injected into
it through a small polypropylene window. To minimize adjustments with
the operational system, a relatively slow beam is generated with
optics on the LO plate, which is refocused with a mirror
attached to the array unit before it enters the meander line. To
provide the correct polarization, two mirrors are used on the LO
plate. A small relay consisting of flat and off-axis paraboloid mirrors
outside the cryostat and a further small paraboloid mirror inside the
cryostat carries the beam from the horn on the LO plate to the LO
coupler. The adjustment to bring the LO beam into alignment is
accomplished using pusher screws to tilt and shift the two
mirrors. The LO coupler is machined from
aluminium. The stop is a 12~mm hole in an absorbing plate
which is used to clean up the beam (Fig.~\ref{locoupler}). Diagnostics are provided by observing the pumped I/V curves on
the mixers to obtain as uniform a distribution of LO power as
possible. An image of the I/V and P(IF)/V curves displayed by the HARP control
software during operation is shown in Fig.~\ref{ivcurves}. 

\subsubsection{Radial probe mixers}

\begin{figure}
\centering
\noindent \includegraphics[width=9cm,angle=0]{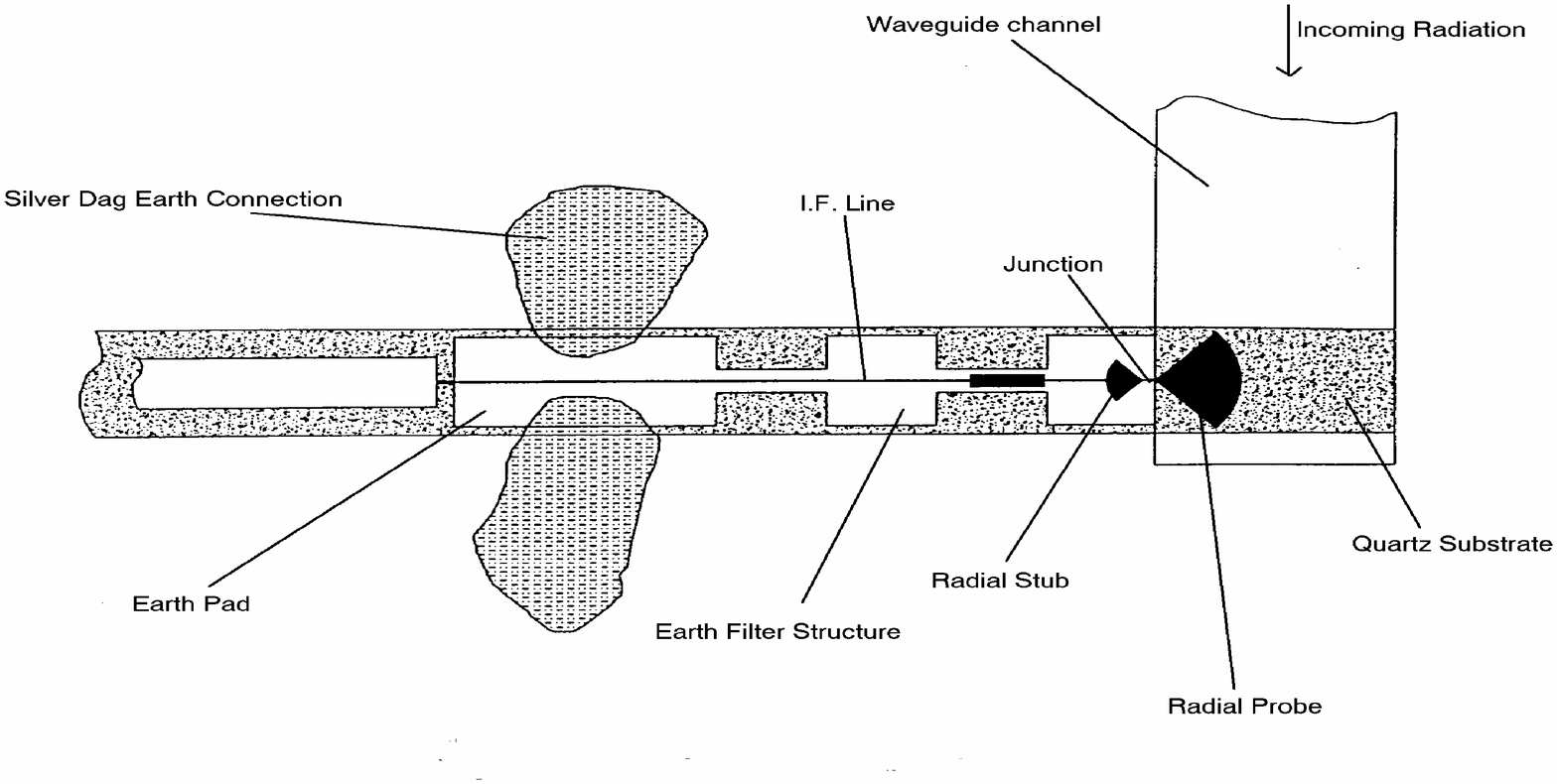}
\caption{\label{mixer} Overview of the radial probe SIS mixer devices. }
\end{figure}

  The radial-probe SIS mixers were designed and developed at the
  Cavendish Astrophysics Group \citep{leech,withington} and
  fabricated in the Kavli Institute of Nanoscience at Delft.  The design, shown in Fig.~\ref{mixer},
  gives extremely broadband operation with similar characteristics for
  each device. A single-sided radial probe couples power from the
  rectangular waveguide to the SIS junction. These radial probes have
  been shown, using scale model experiments and modelling, to
  give a good impedance match to typical SIS junctions \citep{leech,kooi}. The
  capacitance of the SIS junction is tuned out using an inductive
  microstrip stub end loaded with a radial stub to present a short
  circuit at the end of the microstrip.

\begin{figure}
\centering
\noindent \includegraphics[width=8cm]{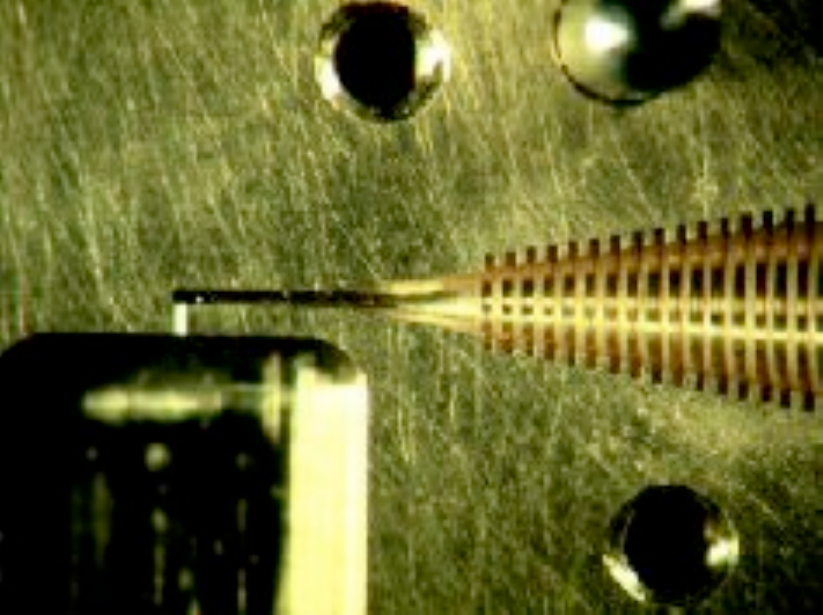}
\caption{\label{horn} Photograph of the HARP horn near the throat, demonstrating the excellent machining quality. The grooves on the horn have a width of 0.1$\pm$0.01~mm.}
\end{figure}

 The imaging receptor is based on the horn-reflector antenna --
 each mixer consists of a corrugated waveguide horn with a reflector
 mirror at its aperture. The feed-horn, waveguide, device and IF slots
 which form the mixer were manufactured as split aluminium blocks,
 then gold-sputtered. This gives blocks that are lightweight, low
 thermal mass, free from waveguide flanges and straight-forward to
 machine. A photograph of one of the finished HARP horns is shown in
 Fig.~\ref{horn}.

  One problem that can arise is the unwanted movement of trapped
  magnetic flux near the SIS junctions.  Typically the unwanted
  Josephson tunnelling of Cooper pairs is suppressed using a coil to
  apply a specific strength of magnetic field across the SIS
  junction. Unfortunately, quantized vortices of magnetic flux,
  usually trapped by defects in the superconducting film near the SIS
  junction, can move around under the influence of external
  electromagnetic fields. If these vortices are close to the SIS
  junction, this can result in a change of the magnetic field applied
  to it. The magnetic field is then no longer at the value needed for
  supercurrent suppression and Cooper pairs begin to tunnel, leading to
  increased noise and decreased stability in the mixer. With large
  arrays, the movement of trapped flux is potentially a major
  problem, and one of the reasons why mixers require different amounts
  of LO power to achieve optimum performance. In HARP,
  the mixers feature arrays of tiny holes in the niobium (Nb) film near the
  SIS junction to prevent the movement of trapped
  flux \citep{leech}. If trapped flux is still proving problematic,
  small resistive heaters can be used to briefly raise the array above
  the critical temperature for Nb, destroying the trapped flux
  vortices.  As a routine start up procedure, the telescope operator
  runs the `all heaters' command (top left of the image in
  Fig.~\ref{ivcurves}), which briefly warms the array. While this
  process is occurring, the changes to the I/V and P(IF)/V curves can be
  monitored in real time. For more localized problems, individual
  heaters can be used on 4 of the detectors at a time. The control
  software screen shown in Fig.~\ref{ivcurves} is thus an extremely
  useful diagnostic tool, and can be used to check the status and
  stability of the HARP mixers throughout an observing run.

\subsubsection{IF system}

The IF system for each receptor consists of a bias tee, isolator and cold
IF amplifier connected to the 4~K stage, and two warm IF amplifiers
separated by a bandpass filter and a fixed level setting attenuator
mounted outside of the cryostat (Fig.~\ref{array}). The co-axial interconnecting cable
from the 4~K stage to the wall of the cryostat is made of stainless
steel to minimize the heat load on the 4~K stage. An IF frequency of
5~GHz, with a minimum bandwidth of 1.6~GHz, is used. The overall
mid-band gain is a minimum of 57~dB, with a noise temperature
$\sim$7~K.

\section{ACSIS and OCS design concepts}
\label{sec:acsis}

\begin{figure*}
\centering
\noindent \includegraphics[width=8cm,angle=-90]{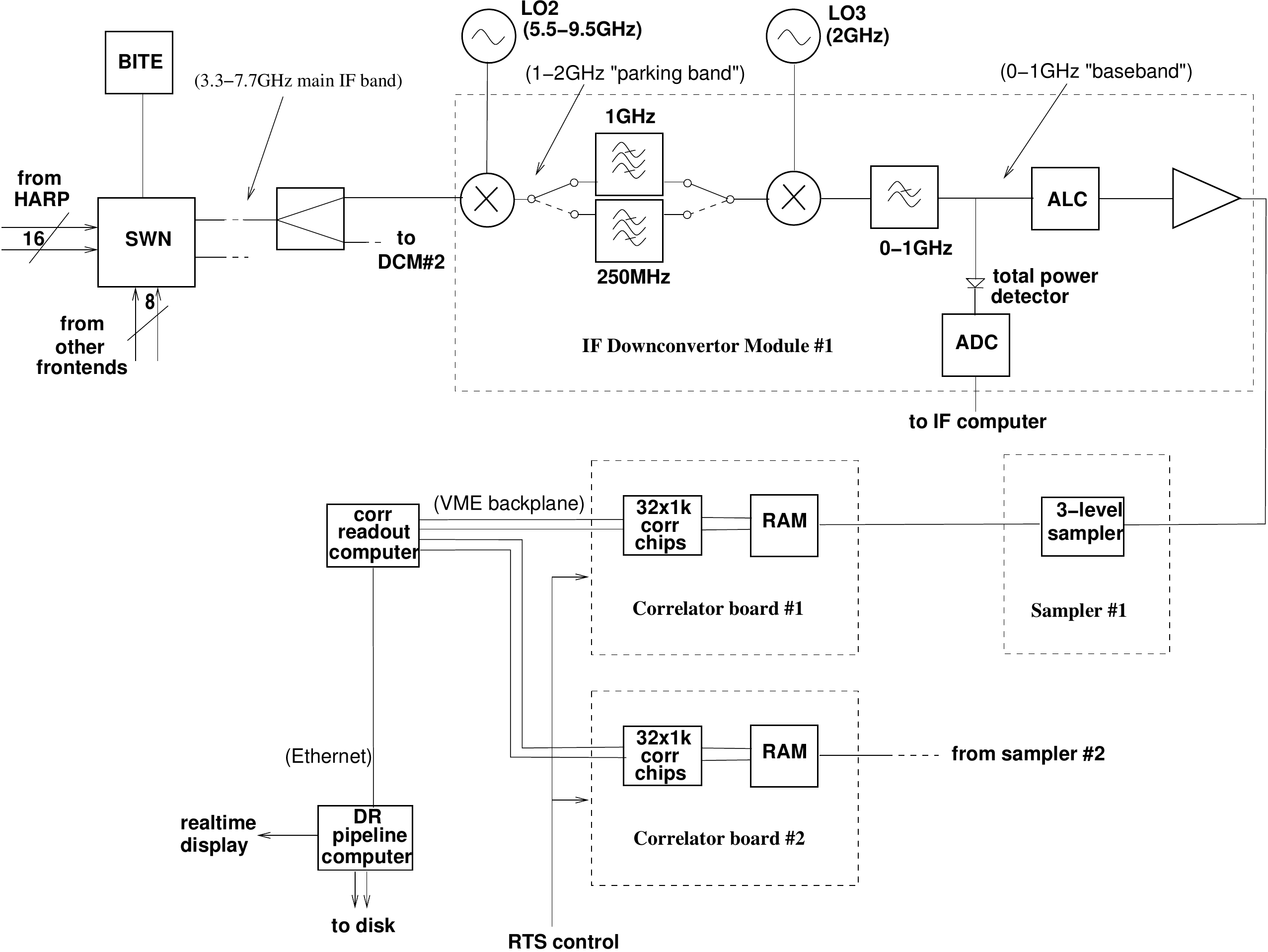}
\caption{\label{fig-acsis1} ACSIS overall system hardware block diagram. Only
one of the 32 Down Convertor Sampler subsections is shown. The signal path is 
described in the text.}
\end{figure*}

ACSIS consists of IF components and samplers, as well as the correlator hardware
and software. ACSIS also includes the initial parts of the
data reduction pipeline and the real time data monitoring facility. The signal path is 
shown in Fig.~\ref{fig-acsis1}, and described in Sec.~\ref{sec:acsisif} and Sec.~\ref{sec:acsis2}. The data storage is described elsewhere \citep{hoveya}. 
The overall control
of heterodyne observations at the observatory, including HARP, ACSIS and
the telescope subsystems, uses real time hardware and
software sequencers \citep{hoveyb}, and is described in Sec.~3.4.
In Sec.~3.1--3.3 below, we describe the ACSIS hardware and software.

\subsection{IF system}
\label{sec:acsisif}

A block diagram showing the IF signal path from the switching network (SWN) is given in Fig.~\ref{fig-acsis1}. The input IF frequencies cover from 3.3 to 7.7~GHz. There are 16 IF signals feeding from HARP and 8 feeds from additional receivers. Switching between these inputs is done by an IF switch. In addition, a Build In Test Unit (BITE) can also feed white noise (with a 3~dB switchable level) and an optional frequency comb to all ACSIS input channels. Each of the 16 HARP IF inputs can feed at least two Down Converter Modules (DCMs) while each of the 8 non-HARP IF inputs, or equally 8 of the HARP IF inputs, can feed 2 or 4 DCMs. Each DCM extracts a nominal 1~GHz or 0.25~GHz wide band from the 3.3 - 7.7~GHz ACSIS IF band. The frequency range extracted is determined by setting the appropriate second LO (LO2). Four tunable LO2s are fed to each set of four DCMs that can be connected to an IF input. A total of 32 DCMs are available. These 2 or 4 DCMs can be placed anywhere in the ACSIS IF band using the four LO2s. Due to the sharing of LO2s the positioning must be the same in all IF inputs.

As described in Sec.~\ref{sec:acsis2} ACSIS can combine the correlators attached to adjacent DCMs doubling the number of frequency channels while halving the effective numbers of available DCMs. These basic modes are listed in Table~\ref{tab-acsis1}. 

\begin{table*}
\begin{minipage}{125mm}
\centering
\caption{ACSIS basic modes.}
\label{tab-acsis1}
\begin{tabular}[t]{rrrrr} \hline
Nominal bandwidth & Usable bandwidth& DCMs &Spectral channels& Channel spacing \\
&/MHz& &&/kHz\\
\hline
250~MHz & $\sim$220 & 1 & 4096  & 61.0 \\
1~GHz   & $\sim$930 & 1 & 1024 & 976.6 \\
250~MHz & $\sim$220 & 2 & 8192  & 30.5 \\
1~GHz   & $\sim$930 & 2 & 2048 & 488.3 \\
\hline
\end{tabular}
\end{minipage}
\end{table*}

With the restriction that all IF inputs are configured identically we can now pick modes until all 32 DCMs are used. For instance for HARP we can chose two 1 GHz bands with 976~kHz resolution or one 1 GHz band with 488.3 kHz resolution. We can also chose one 250 MHz and one 1~GHz with the lower resolution. These setups can be used to observe two lines simultaneously e.g. the $J=3\rightarrow2$ lines of $^{13}$CO and C$^{18}$O - see Sec.~\ref{sec:obs:per}. For non-HARP receivers with a maximum of 8 IF inputs we can chose (i) up to four 250~MHz; or (ii) up to four 1~GHz low resolution bands; or (iii) two high resolution bands; or (iv) two low resolution bands and one high resolution band. There are a large number of combinations possible.

Further, the observing preparation software can place these bands so they overlap in frequency space. The spectra are reduced and stored separately but are later merged by the data reduction software forming hybrid bands. The possibilities including the hybrid modes for HARP are summarized in Table~\ref{tab-acsis2}. In principle, four 1~GHz bands could be combined to form a single 3.7~GHz hybrid spectra for non-HARP receivers, although the current JCMT receivers do not have a usable IF bandwidth as large as this.

\begin{table*}
\begin{minipage}{145mm}
\centering
\caption{ACSIS bandwidth/resolution configurations allowable with HARP.}
\label{tab-acsis2}
\begin{tabular}[t]{rrclrl} \hline
Nominal bandwidth & Usable bandwidth& No. of  & Spectral channels& Channel spacing& Notes \\
&/MHz&sub-bands& per sub-band&/kHz&\\
\hline
250~MHz & $\sim$220 & 1 & 8192  & 30.5 & \\
500~MHz & $\sim$440 & 1 & 8192  & 61.0 & Hybrid configuration \\
1~GHz   & $\sim$930 & 1 & 2048 & 488.3 & \\
2~GHz  & $\sim$1860 & 1 & 2048 & 976.6 & Hybrid configuration \\
250~MHz & $\sim$220 & 2 & 4096  & 61.0 & \\
1~GHz   & $\sim$930 & 2 & 1024 & 976.6 & \\
\hline
\end{tabular}
\end{minipage}
\end{table*}

In the DCM, each sub-band is amplified, filtered and converted to the sampler
frequency range (baseband). The DCMs mix the IF with a tunable LO
(LO2). A final LO (LO3) fixed at
2.0~GHz does the final downconversion to the baseband. 
Rather than having multiple system bandwidth/resolution combinations
using many different IF filters, the philosophy is to have only two
hardware options, and ensure that these cover most astronomical
requirements. Each DCM can therefore switch between
a wideband (1~GHz) and a narrowband (250~MHz) mode (see Table~\ref{tab-acsis1}).
In practice, the edges of the DCM filtering limits the
bandwidth of each DCM to 930~MHz and 220~MHz respectively -- these are the
worst case, the -3~dB power points of the DCMs. After anti-alias
filtering, the total power (TP) signal is measured and the
RF signal is then fed through
an automatic level control (ALC) circuit. The ALC is used to maintain constant
TP into the samplers, in order to ensure the
most stable system bandpass (and the flattest baselines), and the
optimal digitization threshold. The TP signal is recorded and re-applied to the data stream
in the ACSIS real time data reduction.
The DCMs use MMIC architecture and miniaturized components, and plug
into a standard VME backplane. A second section of the backplane includes
SMA ``blind-mate'' connectors for the RF analogue signals, allowing
for simple field replacement of the whole unit.
The IF systems are mounted within temperature-controlled racks, using
a liquid glycol cooler system and external heat exchanger.
On board each DCM unit are temperature
probes and a heater for internal temperature control.
In this way temperatures within the
DCMs are maintained to better than 0.1~K over time-scales of hours. This was done to maintain good passband flatness, particularly in the
hybrid configurations.
A single VME IF control computer is used to read out TP signals and run
the temperature control feedback loop for all 32 DCMs.

\subsection{Samplers and correlators}
\label{sec:acsis2}

Each DCM output is fed to a single sampler unit, running with a fixed clock
speed of 2.0~GHz. In 1~GHz mode this unit digitizes the full bandwidth at the
Nyquist frequency, and in 250~MHz mode it Nyquist samples the 500--750~MHz band,
aliased down to baseband by only keeping each fourth sample.
The sampler uses Sony dual high-speed comparator chips implementing a 3-level sampler. At optimum sampling
levels, their
quantization noise reduces the correlator efficiency to 0.81 (so the noise
in the final spectrum is increased by a factor of 1.23 over a perfect
spectrometer). A serial-to-parallel
converter on each sampler board provides a parallel digital
data stream for connection to the correlator board.

The correlator boards uses the ``QUAINT'' custom correlator
chips \citep{escoffier}, running at a clock speed of 62.5~MHz. The chips were
deliberately run at reduced clock rate to reduce heat
dissipation down to $\sim1.5$~W
(especially important with the reduced atmospheric
pressure on Mauna Kea) in order to increase lifespan.
Each chip has 1024 lags, and 32 chips are mounted on each correlator
board. Digitized data at 2.0~GSamples~s$^{-1}$ is stored rapidly on to
RAM on each board, and read out in parallel
to all 32 correlator chips for
processing at the slower clock rate. Thus each board can provide 1024
lags with 1.0~GHz total bandwidth. In 250 MHz mode the digitized data is stored at 0.5 GSamples~s$^{-1}$ allowing four correlator chips to be chained together creating 8 correlators with 4096 channels each per board.  

The number of channels can further be increased by only recording the incoming digitized data half the time. The time gained allows more correlator chips to be chained together. Two correlator boards time-interleaved in this fashion will correlate all incoming data with twice the number of effective channels. In practice, to avoid extra switching
networks, the IF signal is fed to two DCMs, samplers, and correlator
boards, and the lags synchronously read by the two sets of correlator
boards.

Data is accumulated in the correlator chips, and all 32 chips on each
board can be read out to the real time computers every 100~ms to 10~s.
Data-taking on the correlator boards (and the TP detector)
is controlled through three hardware
control bits fed from the Real Time Sequencer \citep[or RTS; see below and ][]{hoveyb}. Each
data sample has a unique sequence number ($N$), used in the later
data reduction stages.
Each VME real time readout computer is used to read 4 correlator
boards, combine the
autocorrelation functions (ACF) from the chips which are being used in
parallel, and normalize the ACF. The processed data are then
transferred to the main data reduction pipeline computers over the Ethernet connection.

\subsection{Data reduction system}
\label{sec:dr}

Data from all eight real time correlator computers plus the IF computer are fed over
Gigabit Ethernet to the reduction cluster. This cluster consists of eleven
general-purpose
computers which use the DRAMA messaging system \citep*{bailey} for communication,
and run CASA \citep{jaeger} and Glish for the real time data reduction.
Nine of these computers run independent Sync, Reduction and Gridder tasks for
reducing the data. The other two run real time display and
data storage tasks.
The Sync tasks gather relevant data from the
observatory subsystem necessary to describe and reduce each data
sample (with sequence number $N$ -- see Real Time Sequencer
section below -- and DCM number $M$).
Subsystems which provide data to the Sync tasks
include the RTS, TCS, secondary mirror (SMU), beam
rotator, frontend computer, and ACSIS IF and correlator
computers. It has internal buffering to allow different subsystems to
send data asynchronously, and only sends the data package on to the
reduction tasks once all the relevant data are gathered.

The reduction tasks take the data package, multiply the ACF by the IF
TP signal for each sequence number $N$ and DCM channel $M$
($P(N,M)$), Fourier Transform (FT) the ACF and apply a van Vleck
correction \citep{daddario} to give the total power spectrum ($S(N,\nu)$) for each data
sample and DCM channel being used: 
\begin{equation}
\label{eqn1}
$$S(N,M,\nu) = P(N,M) . {\rm FT}({\rm ACF}(N,M))$$
\end{equation}

The reduction task is designed as a stack machine
using object-oriented techniques internally. It is given a recipe as part of the XML sent during the C{\sevensize ONFIGURE} action \citep{lightfoot}. The reduction system creates a real time gridder image for quick-look image and spectral displays, and for focus and pointing calculations.

Depending on the observing mode (see below), the relevant
sky reference power ($S_{\rm off}(M)$) is subtracted and divided out of
each on-source sample power (($S_{\rm on}(N,M)$) to remove the system
bandpass response, and the result calibrated by ${\rm T_{sys}}(M,\nu)$:

\begin{equation}
\label{eqn2}
$$ T_A^*(N,M,\nu)= {T_{sys}(M,\nu).{{S_{\rm on}(N,M,\nu) - S_{\rm off}(M,\nu)}
\over {S_{\rm off}(M,\nu)}}} $$
\end{equation}

The result is a calibrated spectrum for each sample and each
DCM, $T_A^*(N,M,\nu)$, in units of antenna temperature \citep{kutner1981}.
This includes the full header information with sky coordinates and other
relevant data.
These spectral data are saved to disk using a dedicated computer and
task. The maximum
sustainable data rate is limited by the disk access time on this
computer. Currently this is 1-2~MBytes~s$^{-1}$, which is equivalent to a
sample time of 100~ms using HARP in the 1~GHz/2048 channel configuration.

The time-series data are processed using the ORAC Data Reduction Pipeline \citep[ORAC-DR,][]{cavanagh}, creating
    data-cubes from all the observing modes where appropriate, co-adding multiple observations to cover larger areas of
    the sky or to improve signal-to-noise, and performing quality assurance tests. This pipeline runs at the summit and
    also off-line using enhanced data reduction recipes. The pipeline automatically determines baseline regions,
    by analysing the data-cubes and looking for lines using the CUPID  application \citep{berry} and the Starlink KAPPA
    MFITTREND application \citep{currie} and uses this knowledge to calculate high-quality integrated intensity images
    and velocity maps. There are data reduction recipes tuned for narrow-line sources, sources with densely packed lines and also
    broad-line sources. The time-series and processed data are archived at the Canadian Astronomy Data Centre (CADC) and are made
    available through the JCMT Science Archive \citep{economou}.

\subsubsection{Calibration}

In addition to reducing the on-sky data, reduction recipes have been
written to create the $T_{sys}(M,\nu)$ data used for calibration. These
recognize flags on the input stream from the Sync task as being
samples taken on the frontend calibration loads or sky calibration.

\subsection{Control systems - RTS, JOS and the observatory control system}
\label{sec-controlsw}

The real time aspects of data-taking are under control of the RTS.
The RTS is hard-wired to the subsystems throughout the observatory using a
3-wire handshaking system, operating with submillisecond response times.
An observation (for example, a scan map, Sec.~\ref{sec-mlf}) is a recipe
consisting of several
real time sequences of data-taking. During each of the sequences, the RTS
controls the data-taking, i.e. starting and stopping
each integration or sample in the correlator and TP detectors, by asserting and
de-asserting flags over this cable in real time, under the control of a
clock within the RTS. Each individual integration or sample is
assigned a unique sequence number ($N$), which tags the data taken at
that time. The TCS also assigns coordinates for
each sequence number, which allows the off-line reduction system to
determine the coordinates of that data and, for example, to create a
data-cube or map in sky coordinates.
A typical sequence during a scan map
might be a single scan of the telescope across the sky, or an
integration on the sky reference position or calibration load.
In most cases, the start of a sequence in the RTS is triggered by a
telescope `on-source'  flag being asserted; this on-source flag is
one of the inputs to the RTS.

The non time-critical aspects of observatory control, i.e. the setup
of the subsystems before the start of each sequence, is under control
of the JCMT Observation recipe Sequencer \citep[JOS; see
][]{kackley,rees}. This is a high-level control system which executes the
observing recipe, setting up the many sequences within each recipe. It
has a GUI to allow the user to monitor progress and subsystem status
within the recipes, and to start, pause or abort observations.  The
observing recipe is written in Perl, and the subsystem configuration
files are written in XML. The combination of the recipe and the
subsystem configuration files gives a complete description of the
observation.

The control system for HARP is implemented at several different
levels. The JCMT Observatory Control System (OCS) controls HARP
through a series of DRAMA actions that have been standardized for all
frontends at the JCMT. The four key commands are I{\sevensize NITIALIZE}, C{\sevensize ONFIGURE},
S{\sevensize ETUP}\_S{\sevensize EQUENCE}, and S{\sevensize EQUENCE}. The I{\sevensize NITIALIZE} action is used primarily
at the start of an observing session to bring the receiver into a
known passive state. The C{\sevensize ONFIGURE} action tunes the receiver to a new
frequency. S{\sevensize ETUP\_SEQUENCE} prepares the receiver to start an
integration, setting the optics to direct the beam to the sky or to a
particular thermal load. The S{\sevensize EQUENCE} command locks out mechanical
changes to the receiver while the backend is taking data.

The OCS commands to tune HARP are passed to the dedicated HARP
microcomputer, which is a Power-PC based embedded system running the
VxWorks real-time operating system. It receives high-level software
commands from the OCS and other telescope subsystems through the DRAMA
messaging facility over a standard Ethernet connection. The HARP
microcomputer then communicates with the low-level HARP control
electronics via a Controller Area Network (CAN) bus.

A single observation is known as a Minimum Schedulable Block (MSB).
It is written in XML and is a high-level description of the
observation, including the required observing recipe. At the highest
level, the MSB is normally written using the JCMT Observing Tool (OT),
which forms the interface between the user and the system
\citep{folger}. The MSB is a time-independent and system-independent
description of the observation -- i.e. it does not matter when the
observation is actually performed, or how the system is actually
cabled at the time of execution. To convert from the MSB to the real
recipe at runtime, a task known as the translator writes XML
configuration files that the JOS sends to all the subsystems to tell
them the configuration for the observation. The translator also writes
a JOS configuration file (also XML), which tells the JOS the recipe
name, as well as recipe parameters like the step time, minimum number of
integration steps, etc.

\section{Observing modes}
\label{sec-obsmodes}

There are many possible observing modes using ACSIS and the heterodyne
frontends, each of which is a combination of mapping method,
switching method and stepping method. The main mapping methods are
scan, jiggle and stare. The switching methods are: `nod' in
which the telescope is moved, `chop' in which the secondary mirror is moved and
`frequency switch' in which the LO moves.  The stepping methods are `sample' for a single observation and
`grid' for a regular grid or user-defined pattern. Additionally, there are (or are planned) modes which allow skydips,
flat-fielding and calibration.

For HARP, the standard `fiducial' receptor is chosen from one of the four centre receptors, and appears in
the centre of the source when the telescope offsets are zeroed out. This receptor is used for pointing and focus
observations.

\begin{figure}
\centering
\noindent \includegraphics[width=15cm,angle=-90]{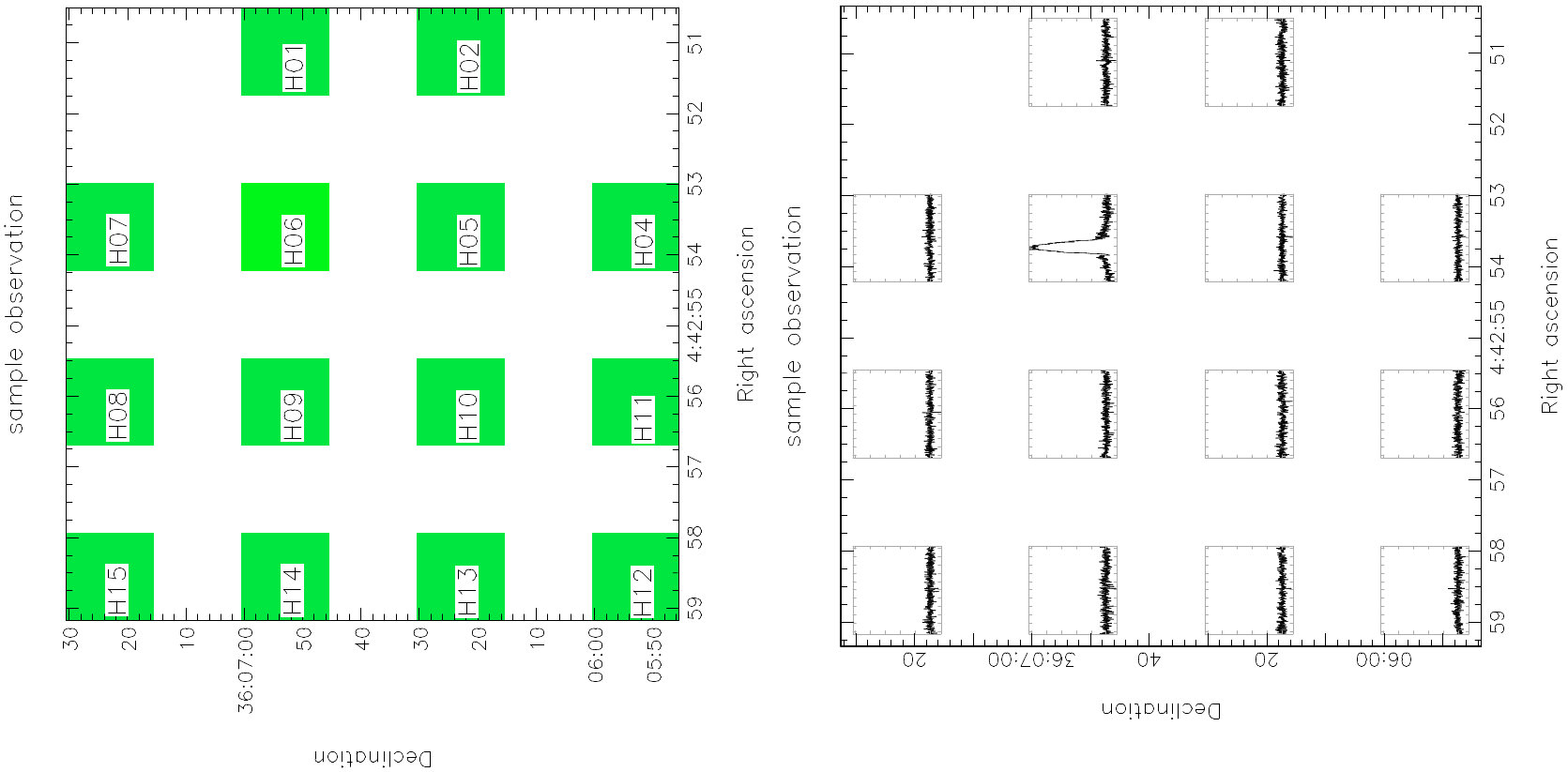}
\caption{\label{stare-pixel} Grid-position-switch stare observation, with receptors labelled. H00 and H03 have been turned off (top). The spectra from each position in the image (bottom). }
\end{figure}

A `stare', or sample observation with HARP is a sparse map covering
$\sim$100~$\times$100~arcsec, with 16 positions spaced at
30~arcsec. The receptors are labelled by the reduction software from
H00 to H15, and the position of each detector on the sky is dependent
upon the orientation of the array, which depends on a combination of
the K-mirror and the source hour angle/elevation. An image of a single
position-switch stare observation, often used for source calibrations,
is shown in Fig.~\ref{stare-pixel} (top) with the receptors labelled. In
this observation, receptors H00 and H03 were turned off, and the pixel
size has been specified in the data reduction as 15~arcsec. In
sample observations, the sky reference position can be obtained using
a position switch, as in this case, or using a chop (also called
beam-switch). The spectra from each position are shown in Fig.~\ref{stare-pixel} (bottom), where emission from the compact source is only seen with detector H06.

Only the highest-priority observing modes have been, or are shortly
expected to be, implemented, and these are described below.

\subsection{Mapping large fields}
\label{sec-mlf}

For mapping large areas, the scanning technique is used. This technique is
also known as on-the-fly mapping, or rastering. This mode requires the highest-performance levels from ACSIS, and is used to produce
fully-sampled rectangular maps of large areas. The most common method
is to cover the whole map as rapidly as possible, and then to co-add
multiple maps to increase the signal-to-noise ratio. This reduces effects
caused by slow sky transmission variations, pointing changes or
calibration drifts. The relative calibration uncertainties across maps
are minimized in this way.

\begin{figure}
\centering
\noindent \includegraphics[width=9cm,angle=-90]{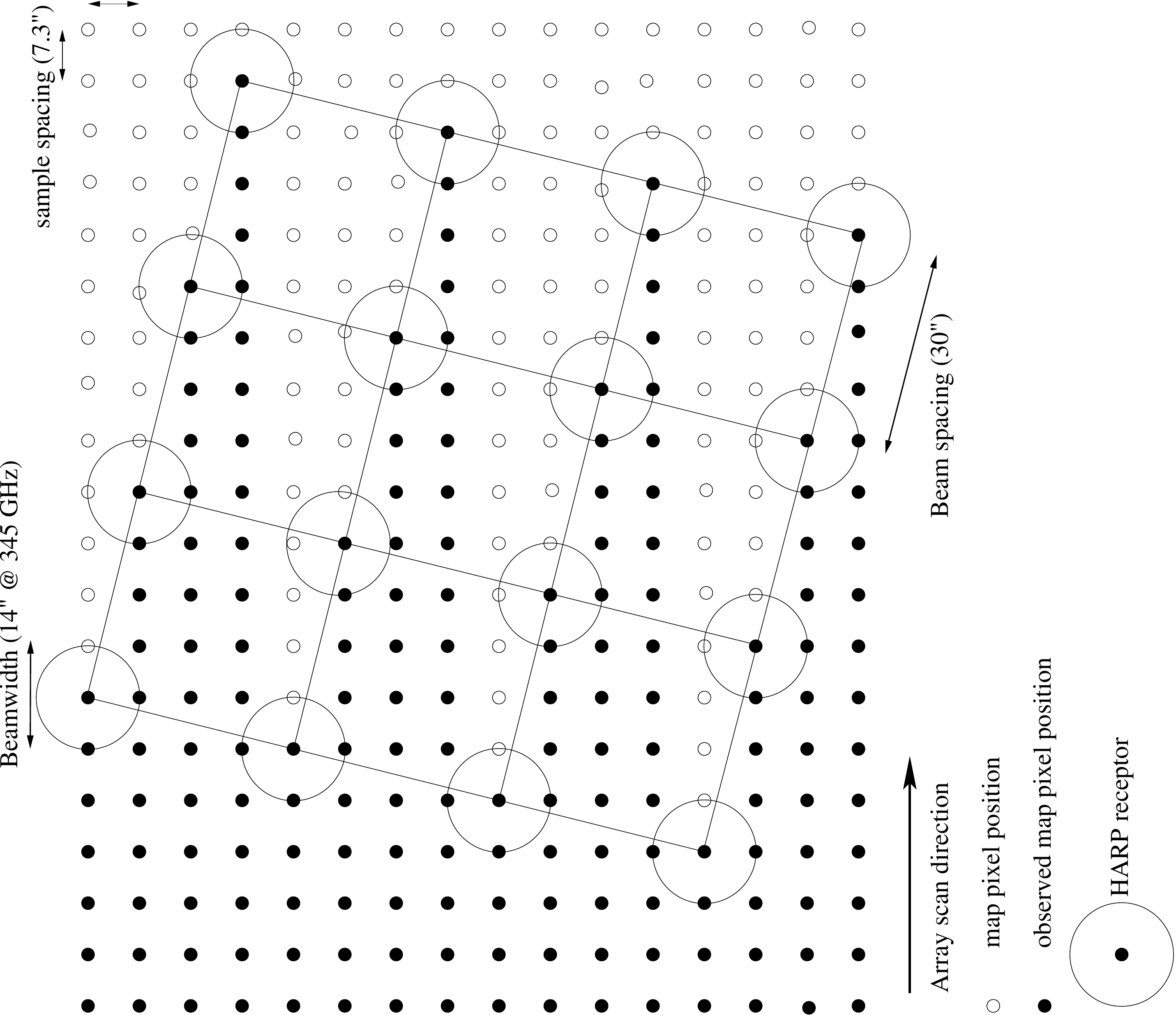}
\caption{\label{raster} Schematic diagram of a scan observation. }
\end{figure}

In the scanning observing mode, the telescope is moved
steadily across the source while the data are integrated and dumped
continuously, up to a maximum rate of once every 100~ms. The K-mirror is used to rotate the array at an angle of 14.48 deg with respect to the scan
direction, which sets the sample spacing perpendicular to the scan direction to be
7.3~arcsec. For HARP/ACSIS at the JCMT, the scan direction reverses for each consecutive row of the scan. Fig.~\ref{raster} is a schematic diagram of a scan
in progress, showing the 16 HARP receptors scanning along a row on a grid of map pixels.

The rectangular map to be observed can be rotated using an area
position angle in the OT, with an appropriate scan position angle so that areas can be observed at any orientation. Effectively, the largest single map that can be made is 1 square degree. The
telescope is nodded to the off-source reference position at the end of
one or more rows, where the default number of rows is specified by the observatory.  The off position is a single, specified reference position, observed in stare mode, which observes simultaneously an off position for each receptor. The off position can be specified in absolute or relative co-ordinates, and can also be specified in a different co-ordinate system than for the target. A
linear combination of the two nearest (in time) applicable off-source
reference observations are used with each individual on-source sample,
and with the nearest (in time) calibration information to form a
fully-calibrated spectrum for each sampling position of the array. The
off-source integration time is a factor $f$ greater than the length of
the individual scanning sample time, so that when combined with the `on'
observations it gives optimum signal-to-noise ratio on the map. $f = t
\sqrt N$, where $N$ is the number of sample points per row, and $t$ is
the sample time. The noise in a single scan does vary due to the change in effective integration time when using an interpolated off.

A basket-weave technique is normally used with scanning modes in order to
minimize the effects of sky and system uncertainties in
a map. This technique scans in different directions for each map, so
that different receptors are used to sample the same sky point, thus
reducing the effects of inaccurate flat-fielding and noisy or non-working
receptors. The optimum change is a 90-degree rotation of the scan
direction, which can be set in the OT.

\begin{figure}
\centering
\noindent \includegraphics[width=7cm,angle=-90]{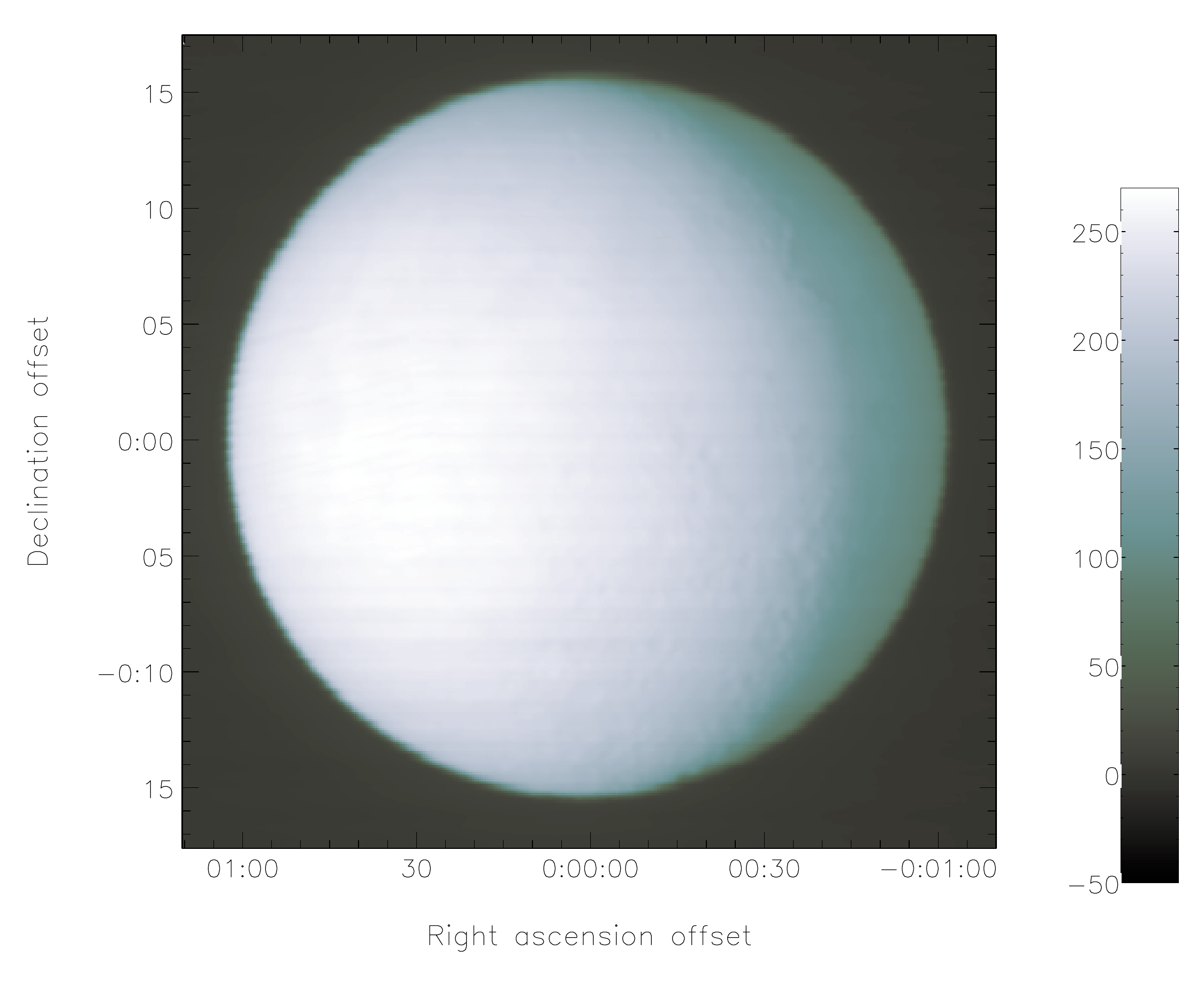}
\caption{\label{moon} Scan observation of a nearly full Moon at 345.79~GHz. }
\end{figure}

Fig.~\ref{moon} shows a 2100~$\times$2100~arcsec scanning
observation of a nearly full Moon at 345.79~GHz, taken during commissioning in December
2006.

\subsection{Mapping small fields}

In order to map regions smaller than 2~$\times$2~arcmin, jiggle or
grid observations can be carried out.

The jiggle-chop (also known as beam-switch) observing mode should be
used for mapping fields smaller than the array field of view where
there are emission-free positions within 180~arcsec of the
source. This is the most efficient observing technique for a fully-sampled map using an array receiver, since it requires only small
movements of the secondary mirror.

In this mode, the secondary mirror is moved to observe a regular
grid-pattern of points on the sky (the jiggle),and then moves off the
source to obtain the sky reference (the chop). The telescope is also
moved to put the source in the opposite beam (a nod) so that half of
the observations are done at a sky position on one side of the source, the other half at
a sky position on the opposite side. This reduces systematics arising from
the secondary chopping to one direction on the sky.

There are two different methods of centring the source within the map. A
receptor-centred jiggle pattern will align the fiducial receptor
with the source, and produce a map that is
asymmetric around the target position. Since the orientation of the
map is determined by the K-mirror rotation, it is not possible to
specify the direction of the asymmetry, and only the central
1.5~$\times$1.5~arcmin is guaranteed to be mapped. This is a
consequence of HARP having a 4$\times$4 grid layout with no central receptor.

\begin{figure}
\centering
\noindent \includegraphics[width=8cm,angle=-90]{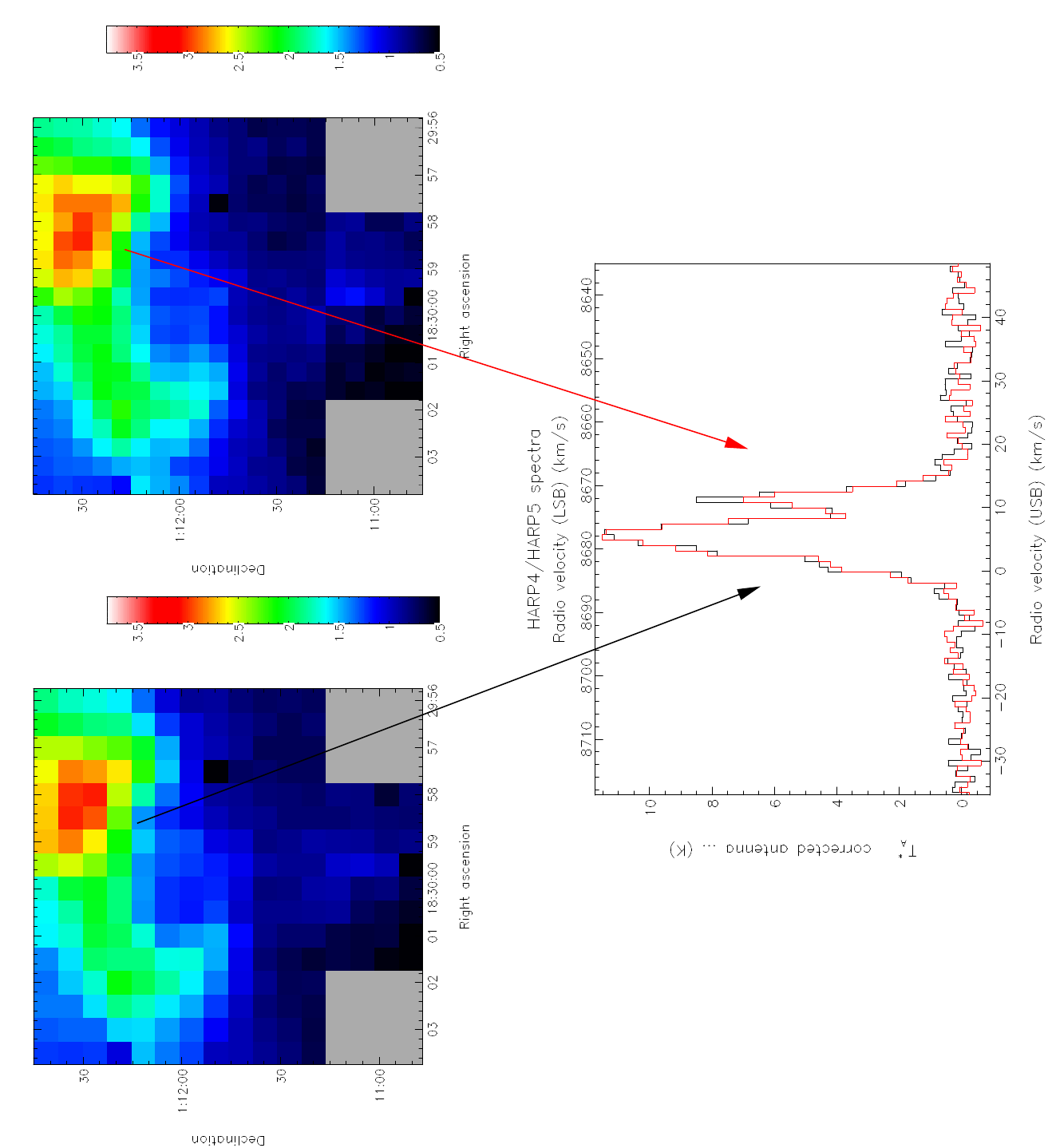}
\caption{\label{jiggles} Integrated intensity maps of jiggle position-switch observations in $^{12}$CO 3$\rightarrow$2 towards a region in Serpens, utilizing the HARP4 pattern (left) and HARP5 pattern (right).  Detectors H00 and H03 have been turned off. Spectra from the same spatial position near the peak are also shown, in a linestyle matching the arrow to the full map (bottom). }
\end{figure}

To fully sample the footprint of the array, special jiggle patterns
called HARP4 and HARP5 have been implemented at the JCMT. HARP4
produces a 16-point rectangular map with 7.5~arcsec spacing
(i.e. slightly under-sampled), while HARP5 produces a 20-point
rectangular map with 6~arcsec spacing, which is slightly
over-sampled with respect to Nyquist sampling of the nominal beam at
345~GHz. Both of the maps will cover a region of
2~$\times$2~arcmin. These two jiggle patterns can be
specified as either pixel-centred, where the source falls in one of
the central four map pixels, creating a slightly asymmetric map; or
map-centred, where the source centre is between the four pixels at the
centre of the map.  HARP4 and HARP5 patterns can be used to fill in the undersampling of the array on the sky, and produce fully-sampled maps. Other jiggle patterns, such as $\rm 3\times3$ or $\rm 11\times11$ can be used to produce under- or over-sampled maps. Fig.~\ref{jiggles} shows HARP4 and HARP5 jiggle observations in
$^{12}$CO $J=3\rightarrow2$ emission of an extended source towards
Serpens, taken during science verification observations for the JCMT
legacy surveys.  Both of these maps show similar spatial and spectral structures, and, as expected, have integrated intensities and standard deviations across the image that differ by less than the variations due to changing observing conditions, such as source elevation and weather. The HARP4 datacube has a lower noise per pixel than the HARP5 datacube.

The jiggle position-switch observing mode is a similar technique,
which can be used for sources where the emission-free region is
relatively far away from the source. In this case, the secondary
mirror is jiggled as described above to observe the source, but the
telescope is moved (a  position-switch) to observe the off position. A grid observing mode is also available. The grid position-switch
moves the telescope to each position in a user supplied grid and executes a stare observation. The grid pattern is supplied in arcsecond offsets, and can
be any shape or size. All of the chop and position-switch modes can be used to carry out sample
observations, in which case only a single point for each receptor is
observed, and a sparse map such as that shown in
Fig.~\ref{stare-pixel} is produced. For all of the jiggle patterns, and for the grid observing mode, either shared off-source reference
positions, defined in the same way as for scans, or
separate off positions for each position in the jiggle pattern can be used. The off position can be specified in the same way as for scan modes. The choice between
these two depends upon the observing mode, the integration time, and how much spatial smoothing will be required in the data analysis. The choice is a compromise between efficiency and photometric accuracy,  since the noise in observations with shared off-source reference positions will be correlated.

\subsection{Frequency switching}

The majority of the observing modes mentioned above will be available
in frequency-switched mode, which is being implemented for jiggle maps in 2009.
In frequency switching with HARP and ACSIS, the switching rate must
generally be faster than the scanning rate. In laboratory
testing, HARP is able to switch at a rate $<$~1~ms, for a step size up to 50~MHz.

\section{System characterisation observations}
\label{sec-sysobs}

\subsection{Beam measurements and efficiencies}

HARP5 jiggle-chop observations of Saturn, with 6~arcsec map pixels, were
carried out in order to determine the beam efficiencies for each
receptor. The observations were made at 345~GHz, and the calculations
were carried out using data on Saturn from the JCMT F{\sevensize LUXES}
programme , with Saturn at a diameter of 19.48 arcsec, and having a brightness temperature of 116.02~K  at the time of the observations. Moon efficiency observations were also carried out at three
frequencies, through sample observations on the full Moon. The Moon
efficiency calculations were made following \cite{mangum}. The results
from these calculations are listed in Table~\ref{tab-eff}. The numbers are  similar to those measured for the previous 350~GHz receiver at the JCMT, RxB3i. Current and archived measurements of efficiencies and receiver temperatures (Sec.~\ref{sec-rc}) are available from the JCMT web site.\footnote{http://www.jach.hawaii.edu/JCMT/spectral\_line/}

\begin{table}
\centering
\caption{\label{tab-eff} HARP efficiencies for Saturn ($\eta _{\rm mb}$) and the Moon ($\eta _{\rm fss}$), measured on 4th/5th February, 2007. }
\begin{tabular}{lrrrr}
\hline
Receptor & $\eta _{\rm mb}$&\multicolumn{3}{c}{ $\eta _{\rm fss}$}\\\cline{3-5}
&345 GHz&330 GHz& 345 GHz& 360 GHz\\\hline
H01	&0.58	&0.70	&0.74	&0.73\\
H04	&0.64	&0.74	&0.78	&0.77\\
H05	&0.64	&0.71	&0.77	&0.76\\
H06	&0.62	&0.71	&0.76	&0.75\\
H07	&0.58	&0.68	&0.75	&0.71\\
H08	&0.58	&0.68	&0.78	&0.75\\
H09	&0.63	&0.72	&0.78	&0.77\\
H10	&0.63	&0.74	&0.78	&0.77\\
H11	&0.58	&0.69	&0.78	&0.76\\
H12	&0.61	&0.73	&0.78	&0.77\\
H13	&0.62	&0.77	&0.79	&0.78\\
H14	&0.58	&0.74	&0.78	&0.77\\
H15	&0.57	&0.74	&0.78	&0.78\\
\hline
\end{tabular}
\end{table}

\subsection{System sensitivity and stability}
\label{sec-rc}

The performance of HARP's receptors exceeds the specification for
receiver temperature ($T_{\rm rx}$) by a large margin. The original specification covered just the central 20~GHz of the tunable range, with the ratio $T_{\rm rx}/\eta_{\rm mb}<$ 330K SSB. Receiver temperatures were
measured using a nitrogen bucket in the receiver cabin in place of the
cold load. Measurements were taken at several local oscillator
frequencies across the tunable band, and are listed for each working
receptor in Table~\ref{tab-rx}. Across the tunable range, the variation in
effective system temperature for each receptor is less than 15~per cent. Current receiver temperature measurements are listed on the JCMT web site.

\begin{table}
\centering
\caption{\label{tab-rx} HARP single sideband receiver temperatures.}
\begin{tabular}{@{}lrrrrr@{}}
\hline
Receptor &\multicolumn{5}{c}{Receiver Temperature (K)}\\
&	328 GHz	&340 GHz&	350 GHz	&360 GHz&	370 GHz\\
\hline
H01&	123&	110&	110&	105&	181\\
H04&	133&	85&	96&	100&	162\\
H05&	165&	82&	93&	87&	127\\
H06&	117&	140&	91&	84&	114\\
H07&	155&	122&	116&	103&	165\\
H08&	148&	93&	93&	97&	147\\
H09&	190&	125&	134&	122&	189\\
H10&	97&	150&	87&	84&	122\\
H11&	120&	145&	105&	92&	143\\
H12&	245&	123&	123&	115&	133\\
H13&	119&	140&	91&	90&	103\\
H14&	99&	144&	95&	86&	116\\
H15&	132&	152&	100&	89&	106\\
\hline
\end{tabular}
\end{table}

\subsection{Sideband rejection}

On-sky observations were made at several different frequencies towards
sources known to have strong emission lines. In most cases, since the
line was not detected in the image sideband, we can only give lower
limits to the rejection. For a few of the detectors, excess noise in
the spectrum meant that the ratio could not be
calculated. Table~\ref{tab-sb} lists the lower limits on sideband
rejection in dB for each detector at several central frequencies. On
average, and across the tuning range, the sideband rejection is better
than 19~dB.

\begin{table}
\centering
\caption{\label{tab-sb}Lower limits on sideband rejection.}
\begin{tabular}{@{}lrrr@{}}
\hline
Receptor &\multicolumn{3}{c}{Rejection (dB)}\\
& 356 GHz&  354 GHz&  345 GHz \\
\hline
H00 &	 &	 &\\	
H01 &	 &	 &	$>$17.86 \\
H02 &	 &	 &\\	
H03 &	 &	 &	\\
H04 &	$>$21.84 &	$>$15.73 &	$>$17.98\\
H05 &	$>$25.09 &	$>$37.31 &	$>$17.77\\
H06 &&	$>$15.88  &	$>$40.48 \\
H07 &	$>$20.56 & &	\\	
H08 &	 &	$>$18.97 &	$>$16.02\\
H09 &	$>$17.00 & 	$>$16.11&	$>$16.39\\
H10 &	$>$30.98 &	$>$29.91&	$>$27.21\\
H11 &	$>$18.25 &&		$>$21.43\\
H12 &	$>$28.61 &	$>$15.03&	$>$24.45\\
H13 &	$>$17.91 &	$>$17.19&	$>$40.66\\
H14 &	$>$22.40 &	$>$19.14&	$>$17.16\\
H15 &	 &	$>$27.02&	$>$20.20\\
\hline
\end{tabular}
\end{table}

\subsection{Calibration}
\label{sec:calibration}
Much work has been done on characterizing and improving the new spectral imaging system. This is a complex task, since the system is composed of a frontend, backend and K-mirror, all of which are newly-commissioned, plus telescope upgrades, and newly-written telescope, data acquisition, and data reduction software. For the integrated systems at the JCMT, under good conditions, and assuming pointing and focus are good, the uncertainties associated with calibrational accuracy of any of the spectral frontends combine to a total of $\sim$~20~per cent. The factors contributing to this are explained in full on the JCMT web page.\footnote{http://docs.jach.hawaii.edu/JCMT/HET/GUIDE/het\_guide/} For the new spectral imaging system, a review of the calibration observations taken for the Gould Belt Survey \citep{ward} show HARP intensity values (mostly) below the reported JCMT standard values by up to 23~per cent, with a mean difference of 15~per cent, for calibration observations taken at 345~GHz, 329~GHz and 330~GHz.

\section{Scientific observations}
\label{sec-sc-obs}

HARP and ACSIS provide simultaneous mapping and high-resolution spectroscopy with high sensitivity in the 325 to 375~GHz (850~$\umu$m) band. The spatial dimension of these maps is well matched to continuum images obtained with SCUBA, and which will be obtained with SCUBA-2. The spectral dimension of the HARP/ACSIS maps provides far more information than can be obtained with continuum images. In this section, we describe some of the first science observations taken with the new system as part of the HARP/ACSIS instrument teams guaranteed time.

\begin{figure}
\centering
\noindent \includegraphics[width=8cm]{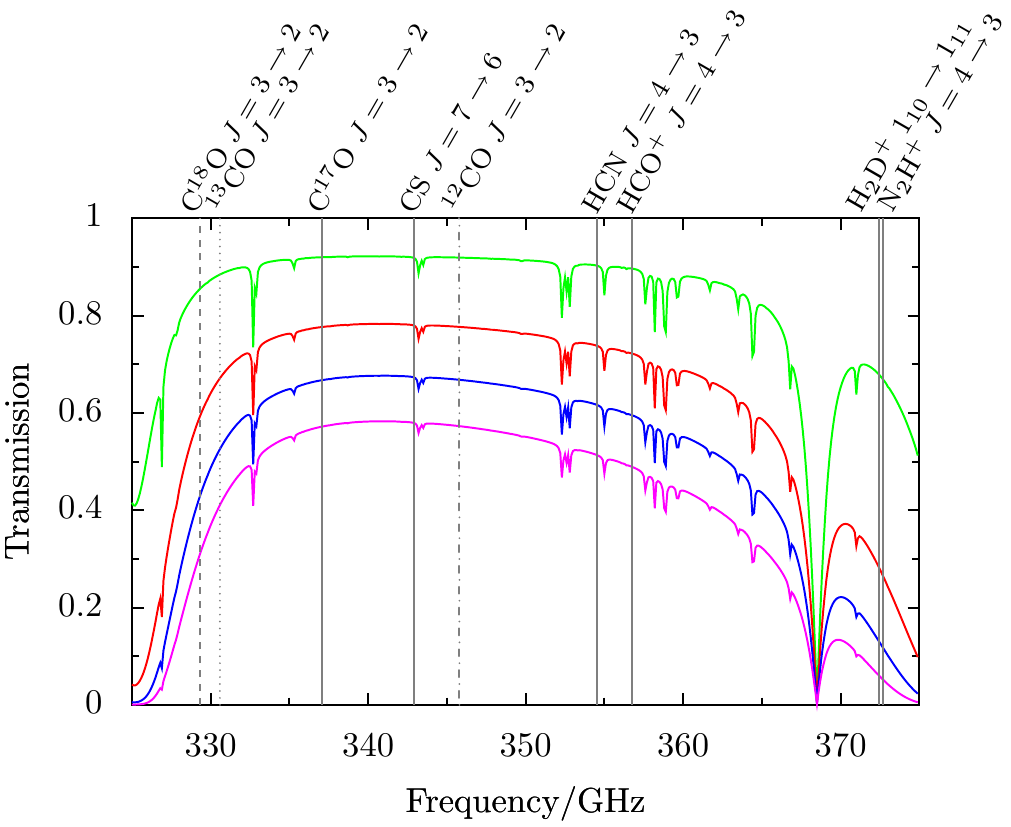}
\caption{\label{fig-atmt} Atmospheric transmission for several different amounts of water vapour at the JCMT in the HARP observing range, corresponding to JCMT weather bands 1 to 4. Astrophysically important lines are marked. }
\end{figure}

\begin{figure}
\centering
\noindent \includegraphics[width=7cm,angle=-90]{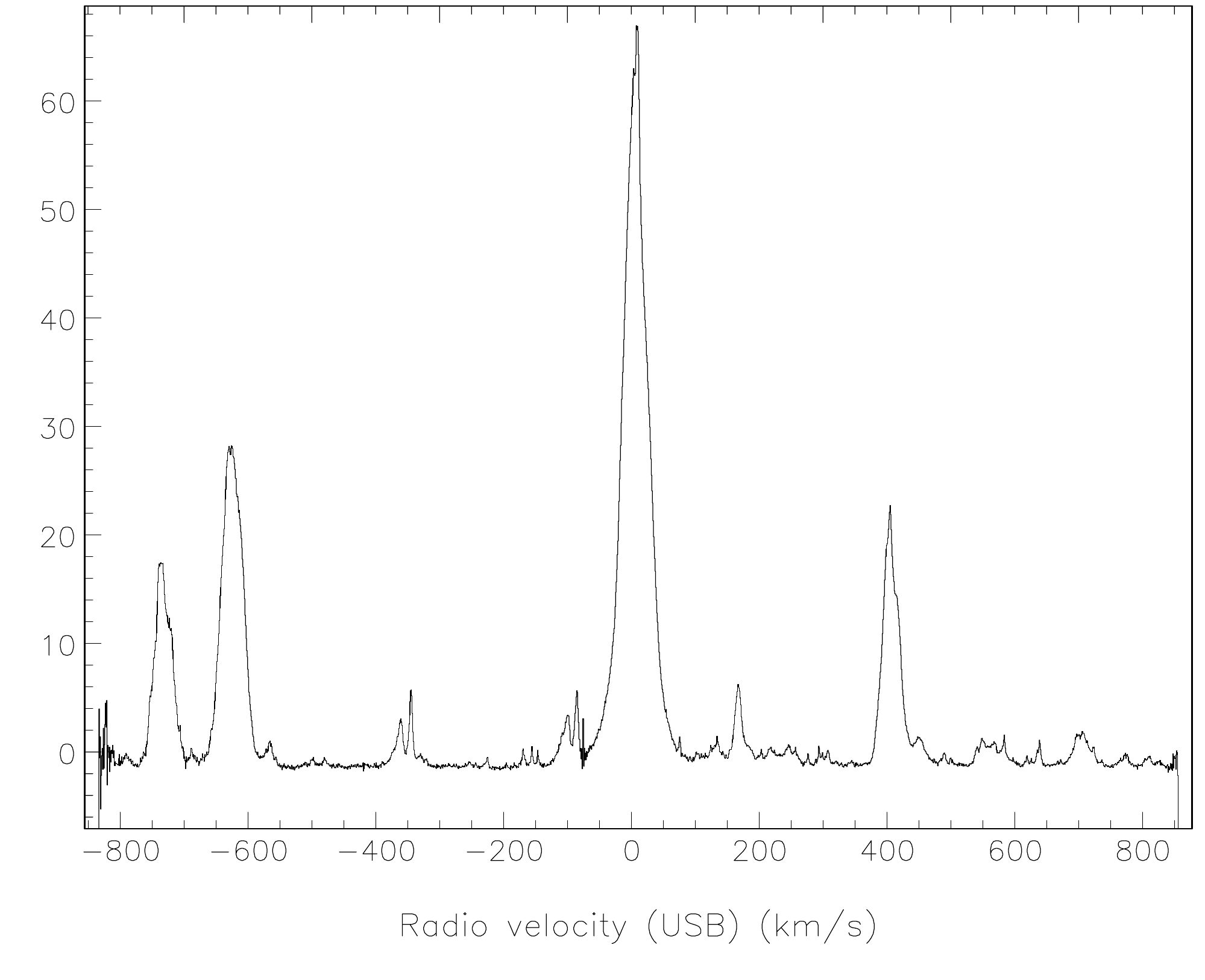}
\caption{\label{fig-2g} Wideband spectrum ($\Delta \nu$=1.9~GHz) at 345 GHz towards OMC1, showing the wealth of lines observable in this frequency range, observable in a single tuning. }
\end{figure}

Comparisons of different lines enables the temperature, space density and composition of the gas to be derived, all of which are essential in understanding the underlying astrophysical processes.  The molecular transitions available in the HARP frequency range trace moderate density (10$^3$ to 10$^7~{\rm cm}^{-3}$) and temperature (10 to 100~K) regimes, probing the warmer, more excited gas associated with energetic events such as star formation.  With the instrument placed at the JCMT on Mauna Kea, the atmospheric transmission allows sufficiently dry nights to obtain good sensitivity at the edges of the band, where astrophysically important transitions from molecules such as C$^{18}$O and N$_2$H$^+$ lie. Fig.~\ref{fig-atmt} shows the position of these lines, along with the atmospheric transmission across the frequency range. Table \ref{tab-lines} lists a sample of the lines observable with HARP, and Fig.~\ref{fig-2g} shows a wideband spectrum at 345~GHz taken towards OMC1, where many lines can be observed.

\begin{table}
\centering
\caption{\label{tab-lines}Some of the important molecular lines observable with HARP. Examples of lines which can be observed simultaneously have been listed in pairs.}
\begin{tabular}{@{}lr@{}}
\hline
Transition &Frequency\\
& /GHz \\
\hline
C$^{18}$O/$^{13}$CO  3$\to$2 & 329.330/330.587	 \\	
 C$^{17}$O 3$\to$2/C$^{34}$S 7$\to$6	& 337.061/337.396\\	
 CH$_3$OH 7$_{14}$ $\to$  6$_{14}$& 338.345	\\
CS 7 $\to$ 6& 342.883	\\
 H$^{13}$CN 4 $\to$ 3& 345.340 	\\
CO 3 $\to$ 2 & 345.795	\\	
 H$^{13}$CO$^+$ 4 $\to$ 3/SiO 8 $\to$ 7& 346.999/347.330\\
HCN 4 $\to$ 3 &354.505	\\
HCO$^+$ 4 $\to$ 3 &356.734	\\
DCN/HNC 4 $\to$ 3 &362.046/362.630	\\
H$_2$CO 5$_{05}$ $\to$ 4$_{04}$ &362.736	\\
H$_2$D$^+$ 1$_{10}\to$1$_{11}$/N$_2$H$^+$ 4 $\to $3  &372.421/372.672	\\
\hline
\end{tabular}
\end{table}

\subsection{Large-scale mapping of molecular clouds}
\subsubsection{Perseus}
\label{sec:obs:per}
\begin{figure*}
\centering
\vbox{
\hbox{
\noindent \includegraphics[width=8cm]{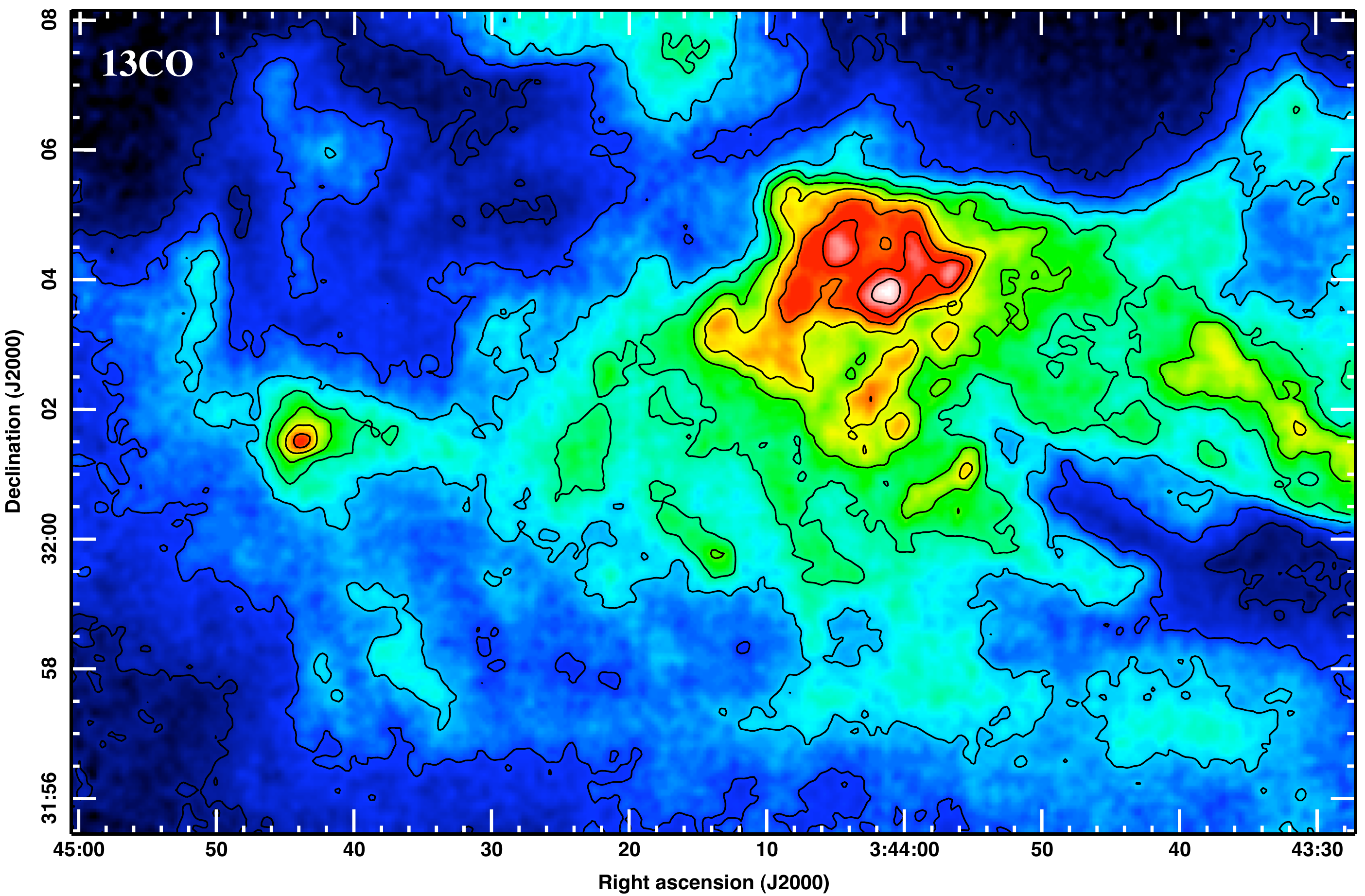}
\noindent \includegraphics[width=8cm]{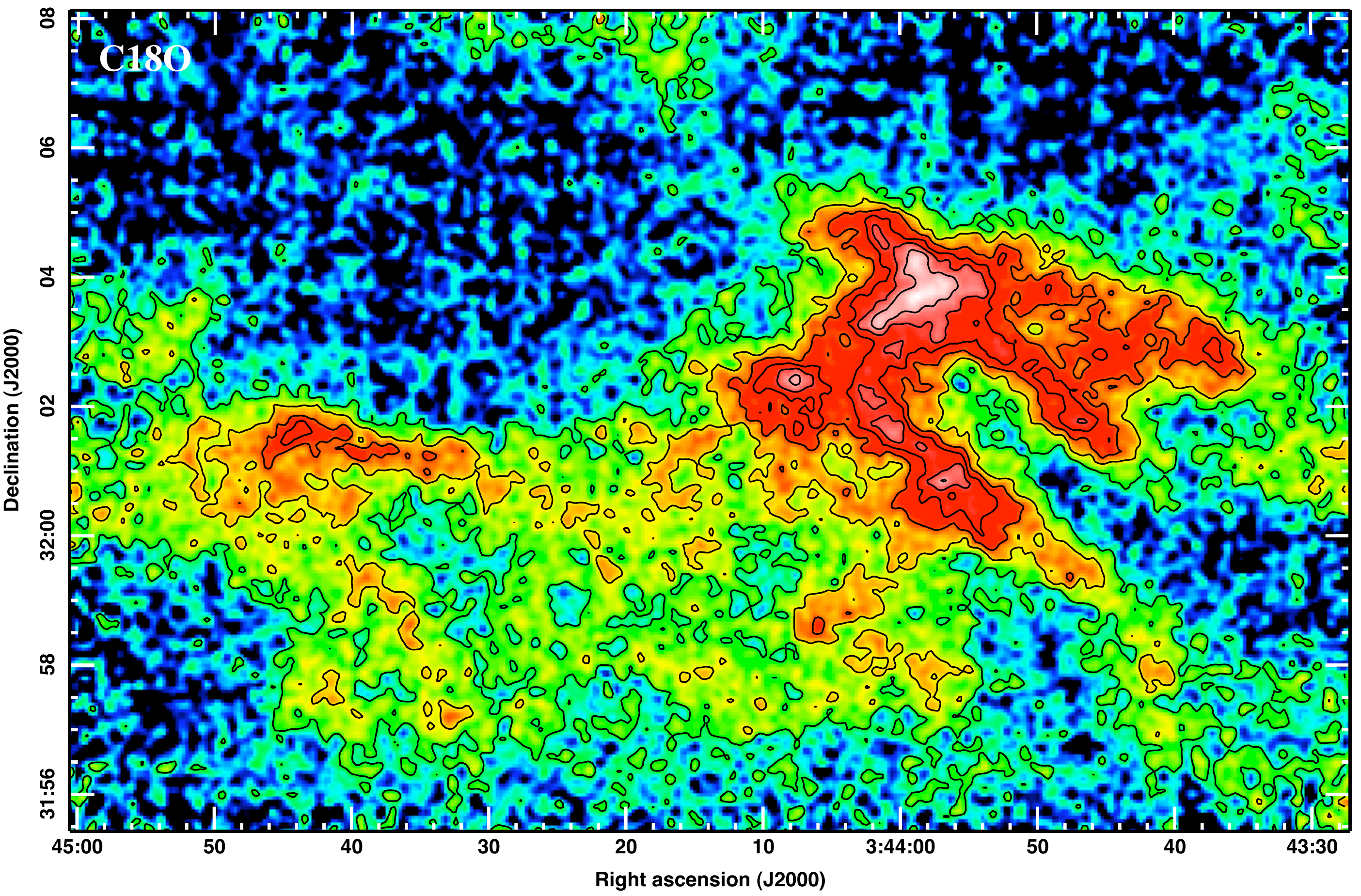}}
\hbox{
\noindent \includegraphics[width=8cm]{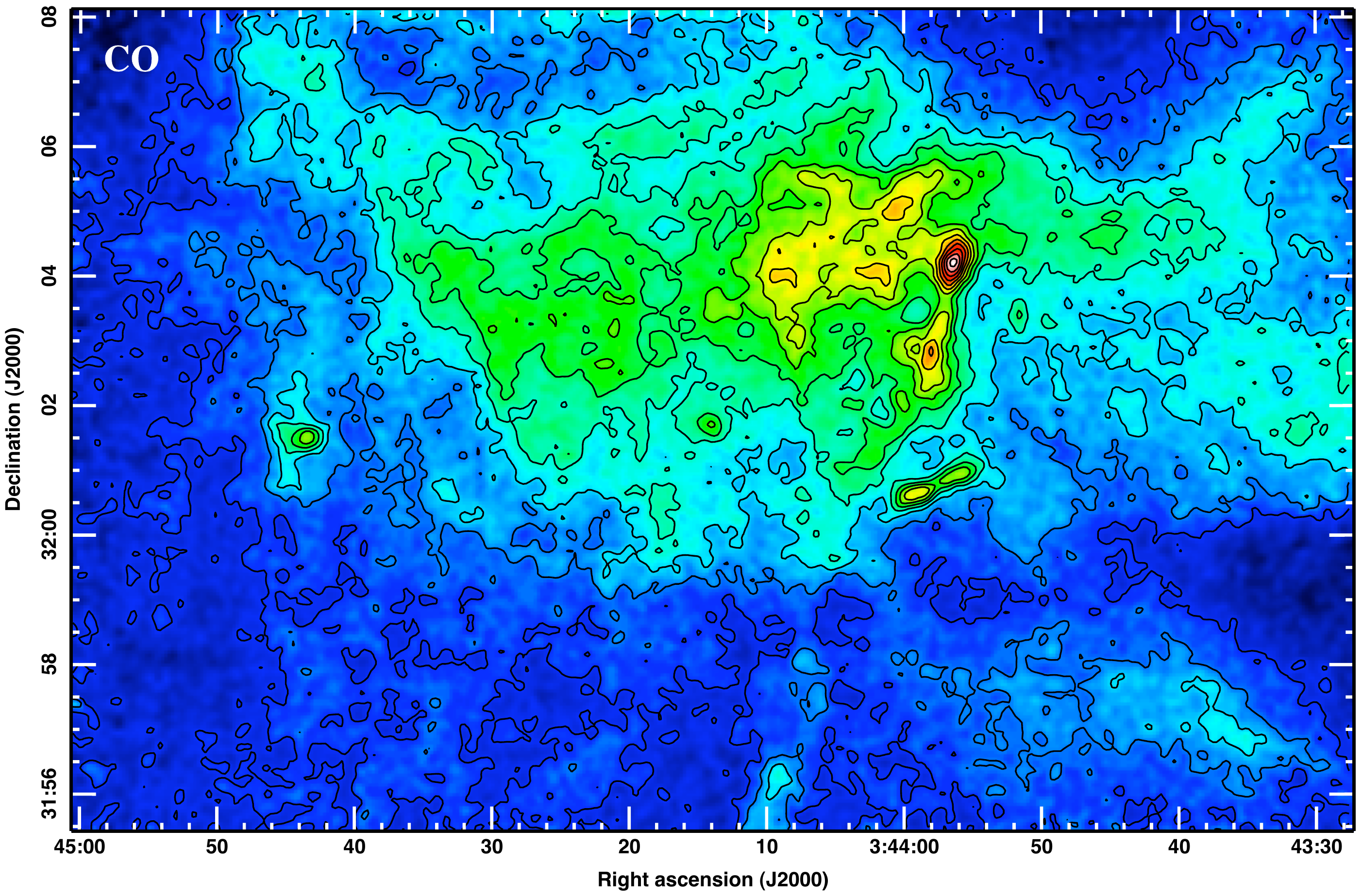}
\noindent \includegraphics[width=8cm]{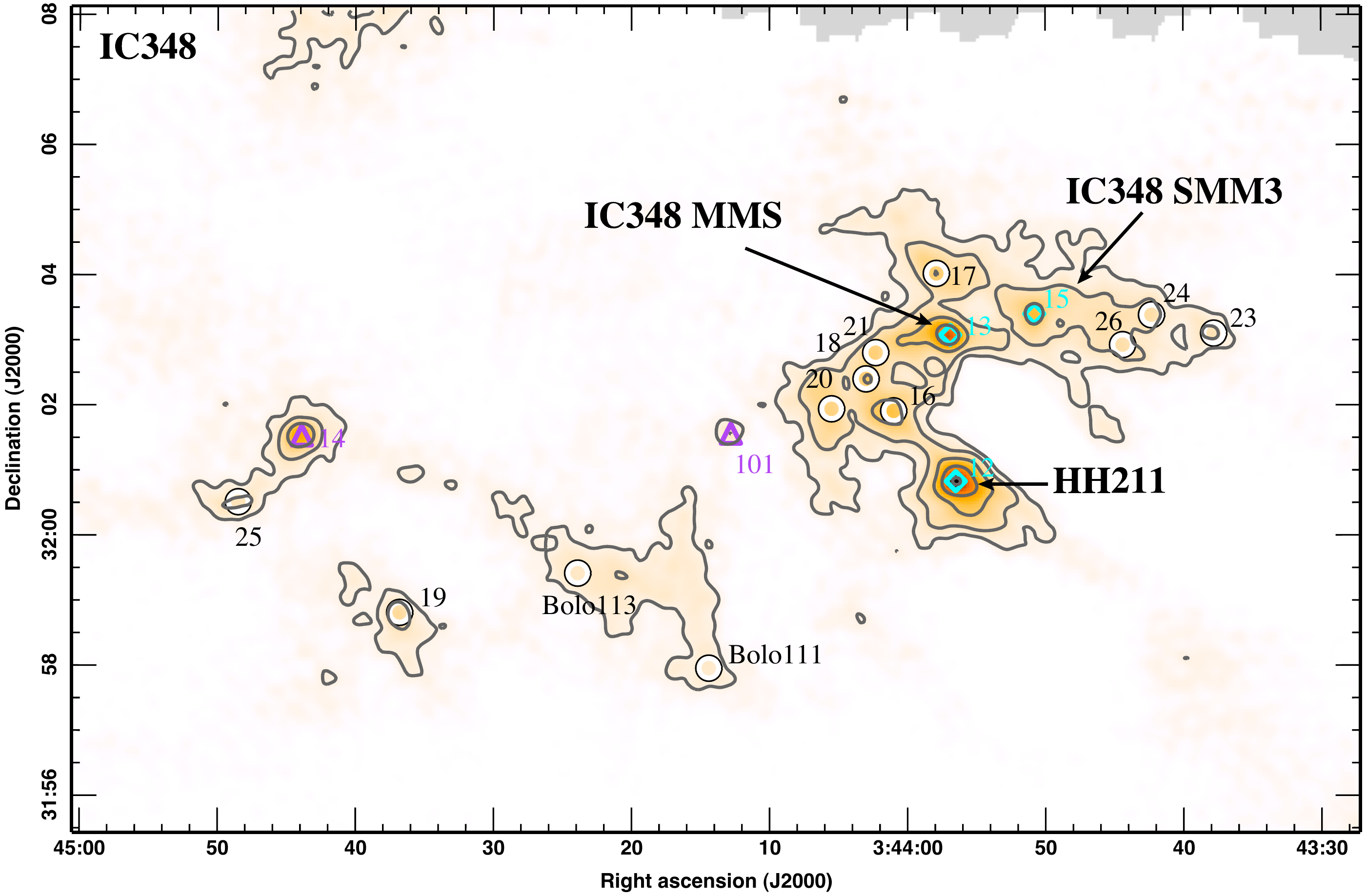}}}
\caption{\label{ic348} HARP $^{13}$CO (top left), C$^{18}$O (top right) and $^{12}$CO (bottom left) $J=3 \to 2$ integrated intensity images towards IC348 taken as part of the guaranteed time. For comparison a SCUBA 850~$\mu$m dust emission map is also show \citep[bottom right,][]{hatchella}, with cores classified by \citet{hatchellb}, and marked as starless (black and white circles), Class 0 (blue diamonds) or Class I protostars (purple triangles). The optically thin C$^{18}$O map closely follows the compact SCUBA emission, although there are some differences, while the $^{13}$CO traces the bulk of the ambient cloud. By contrast the $^{12}$CO map is dominated by the emission from young protostars in the region, notably HH211 (the northwest-southeast oriented 'dumbbell' at 03$^{h}$27$^{m}$39$^{s}$,+30$^{\arcdeg}$13$^{\arcmin}$00$^{\arcsec}$) and IC348-MMS (the nearly north-south oriented flow centred at 03$^{h}$27$^{m}$43$^{s}$, +30$^{\arcdeg}$12$^{\arcmin}$28$^{\arcsec}$)  }
\end{figure*}

Mapping the kinematics of star formation can provide crucial
discriminators between models of star formation. As part of the
HARP guaranteed-time programme of observations we surveyed gas kinematics
in the CO, C$^{18}$O and $^{13}$CO $J = 3 \to 2$ spectral lines across a
wide region of the Perseus molecular cloud \citep*{curtis}. These three tracers are a
powerful diagnostic tool in understanding star-forming regions and
will be extensively utilized in the forthcoming Gould Belt Legacy
Survey \citep{ward}. Of particular importance for the Legacy Survey is the multiple sub-band mode, where C$^{18}$O and $^{13}$CO $3 \to 2$  can be observed simultaneously, and CO can be observed in a high resolution/narrow band mode simultaneously with a medium resolution/wide band mode. One of the mapped regions in the east of Perseus, IC348,
is shown in the different isotopologues in Fig. \ref{ic348}. All of these data took just 12 hours to collect,
compared to the roughly 4 weeks expected with the previous B-band receiver on JCMT. The region
is a dense molecular ridge of about 200~$\msol$, some 10~arcmin to the south-west
of the IR cluster IC348 itself. The centre of the cluster is almost
devoid of star-forming activity \citep{luhman} in contrast to this ridge which
contains numerous sites of low-mass star formation \citep{walawender}. It notably contains the well-studied
extremely collimated young outflow, HH211 (clear in the CO maps as the
dumbbell shaped object, Fig.~\ref{ic348}). Discovered by
\citet*{mccaughrean}, it is the best example of a molecular \emph{jet},
with intense molecular beams terminating in H$_2$ bow-shocks
\cite[e.g.][]{chandler}.  

When used in combination with SCUBA dust continuum maps, the HARP data
offer powerful insights into the velocity structure of the
regions and optical depths. The optically thin C$^{18}$O traces the velocity
structure in and around the dense cores, identified in SCUBA maps
\citep{hatchella,hatchellb}
which can be used to explore the turbulent environment just outside of
the cores. The total intensity traces the dust very accurately
and individual star-forming cores may be differentiated. The $^{13}$CO data probe the bulk motions of the cloud and
trace the highest-density protostellar outflows. Numerous cavities
and clumps are clear, which may be relics of a previous generation of
star-forming objects \citep[e.g.][]{quillen}. Finally, the $^{12}$CO data trace mainly
high-velocity outflows, which may regulate the whole process by feeding-in large quantities of energy and momentum into the cloud bulk
\citep{norman,li}. In
this region three or four outflows are clear, totalling some $1.2
\times 10^{37}$~J of kinetic energy which is close to the cloud's
  turbulent energy, $2.7 \times 10^{37}$~J, suggesting the possible link
  between outflows and cloud turbulence. 

\subsubsection{Orion}
The guaranteed time observations carried out jointly between the HARP
and ACSIS teams observed 2.7~deg$^2$ of the Orion A molecular
cloud in $^{12}$CO $J=$3$\to$2. This region contains the L1641 cloud and
the integral filament (with OMC1), allowing study of both extremely
bright (OMC1 and L1641N) and more quiescent regions (further
south). The scientific analysis being carried out is primarily aimed
at producing a census of protostellar outflows in the region. Currently these
outflows are being identified and characterized with respect to their
energy, momentum and mass, whilst investigating their effect on the cloud
as a whole. Our $^{12}$CO observations allow an estimate of the mass
of the whole cloud, which places a lower limit on the
mass contained in the region observed here of $10^{3}$~\msol. This
accords well with \citet{bally} who found a mass of around
$5\times10^4$~\msol\ in a an area over twice as large (8 deg$^2$ compared with 2.7 deg$^2$) observed in a less optically thick tracer ($^{13}$CO J=1$\rightarrow$0).

\begin{figure}
\centering
%\hbox{
\includegraphics[width=9cm,angle=-90]{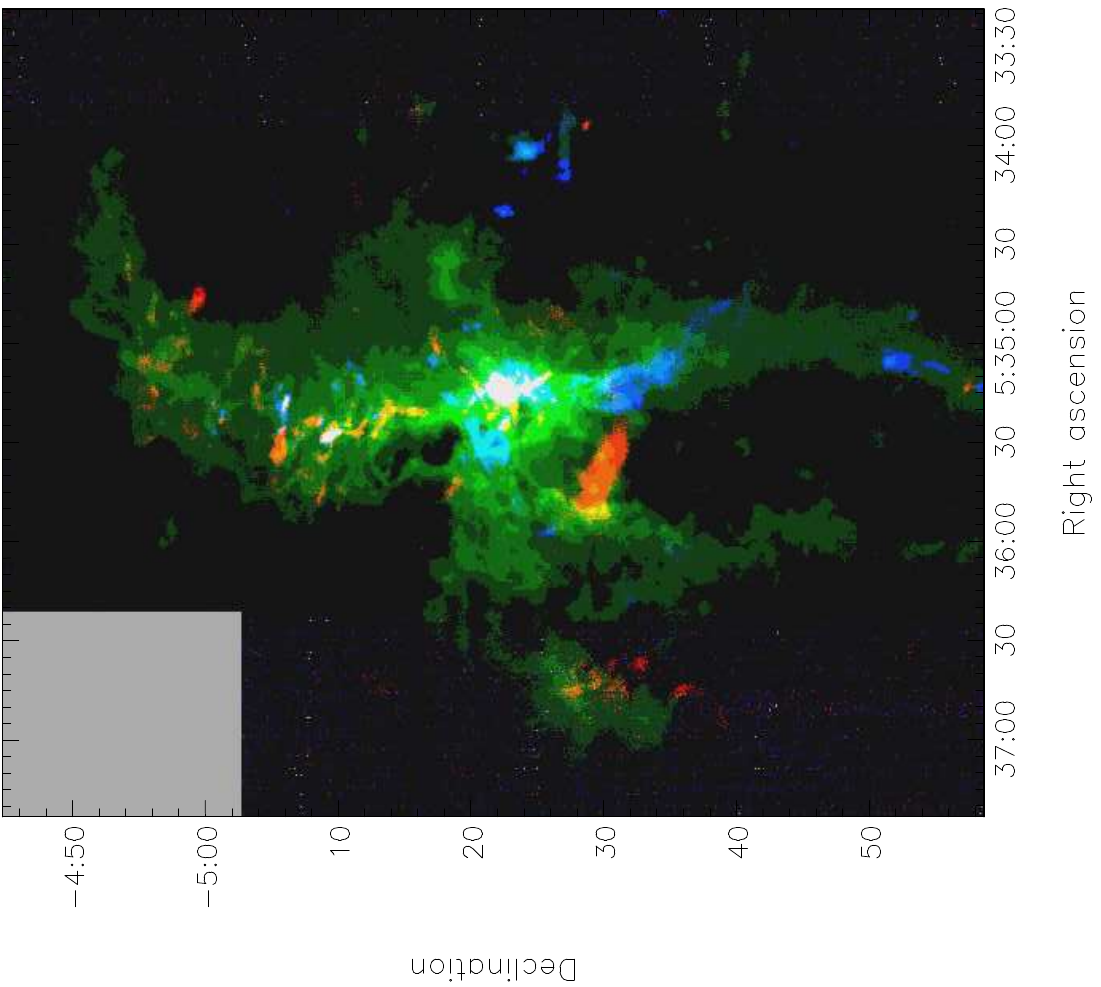}
%}
%\caption{ (Left) The Orion filament, integrated from -0.9 to 17.1~\kms. Red and blue contours show the outflowing gas, integrated from 14.1 to 34.1 and -14.8 to 5.1~\kms. (Right) The same integrated map shown as a red/blue/green false colour image.\label{fig-integfilament}}
\caption{The Orion filament, integrated from -0.9 to 17.1~\kms\ (green), 14.1 to 34.1~\kms\ (red) and -14.8 to 5.1~\kms\ (blue) shown as an rgb false colour image.
 \label{fig-integfilament}}
\end{figure}

\begin{figure}
\centering
\includegraphics[width=8cm]{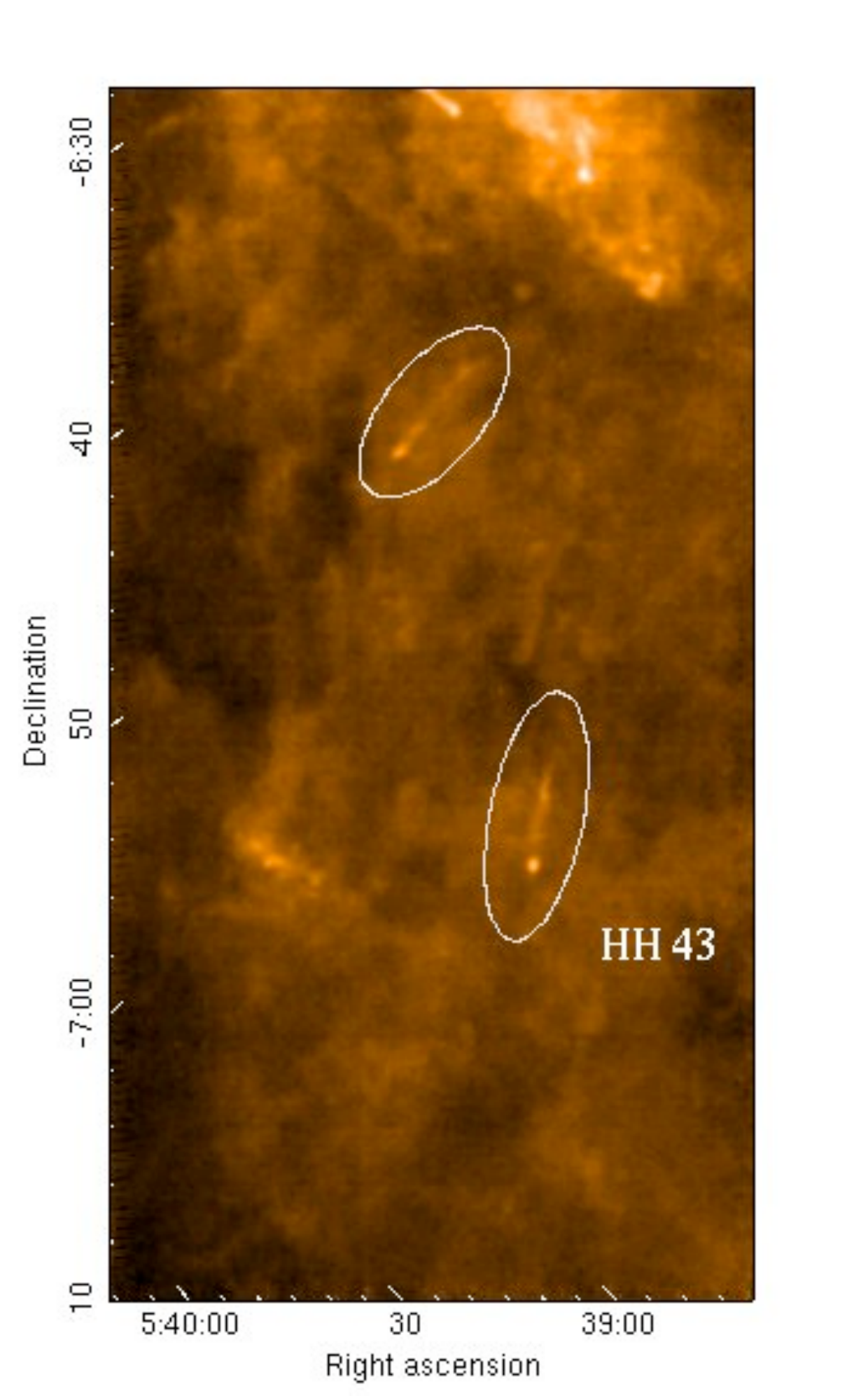}
\caption{A region from the Orion A HARP/ACSIS GT survey. Two
prominent jet-like structures are highlighted. Both contain a bright
knot-like `nose' with a fainter `tail' streaming away. The lower one
is at the position of the Herbig--Haro object HH 43. \label{fig-field3}}
\end{figure}

Fig.~\ref{fig-integfilament} shows some of the strongly red- and blue-shifted gas located in the integral filament as a red-green-blue false-colour image. Fig.~\ref{fig-field3} shows an integrated intensity map from a more
southerly region (below L1641N). Although the gas here is less bright
and more quiescent, Fig. \ref{fig-field3} clearly shows some exciting
jet-like structures. Some of these are associated with HH objects,
such as HH 43 in the current image. The bright clump of emission is at
the position of HH 43 -- which also appears as a strong knot when
observed in infrared 2.12 \micron\ H$_2$ data in \citet*{stankea,stankeb}. The fainter collimated tail extending NW aligns
well with the short flows SMZ68 
\citep[identified as a bow shock in][]{stankeb} and the longer SMZ67, a giant flow identified from
a number of knots between HH 43, 38 and 64, and extending considerably
further than the visible `tail' seen in CO. The giant H$_2$ flow has a length
of 1.45~pc, compared with approximately 0.7~pc, or 5~arcmin in CO. Their
predicted driving source for the giant flow is at a position just at
the NW end of the tail seen in CO.

\subsection{High resolution spectral imaging of protostellar cores}
\label{sec:obs:cores}
\begin{figure*}
\centering
\hbox{
\includegraphics[angle=-90,width=7cm]{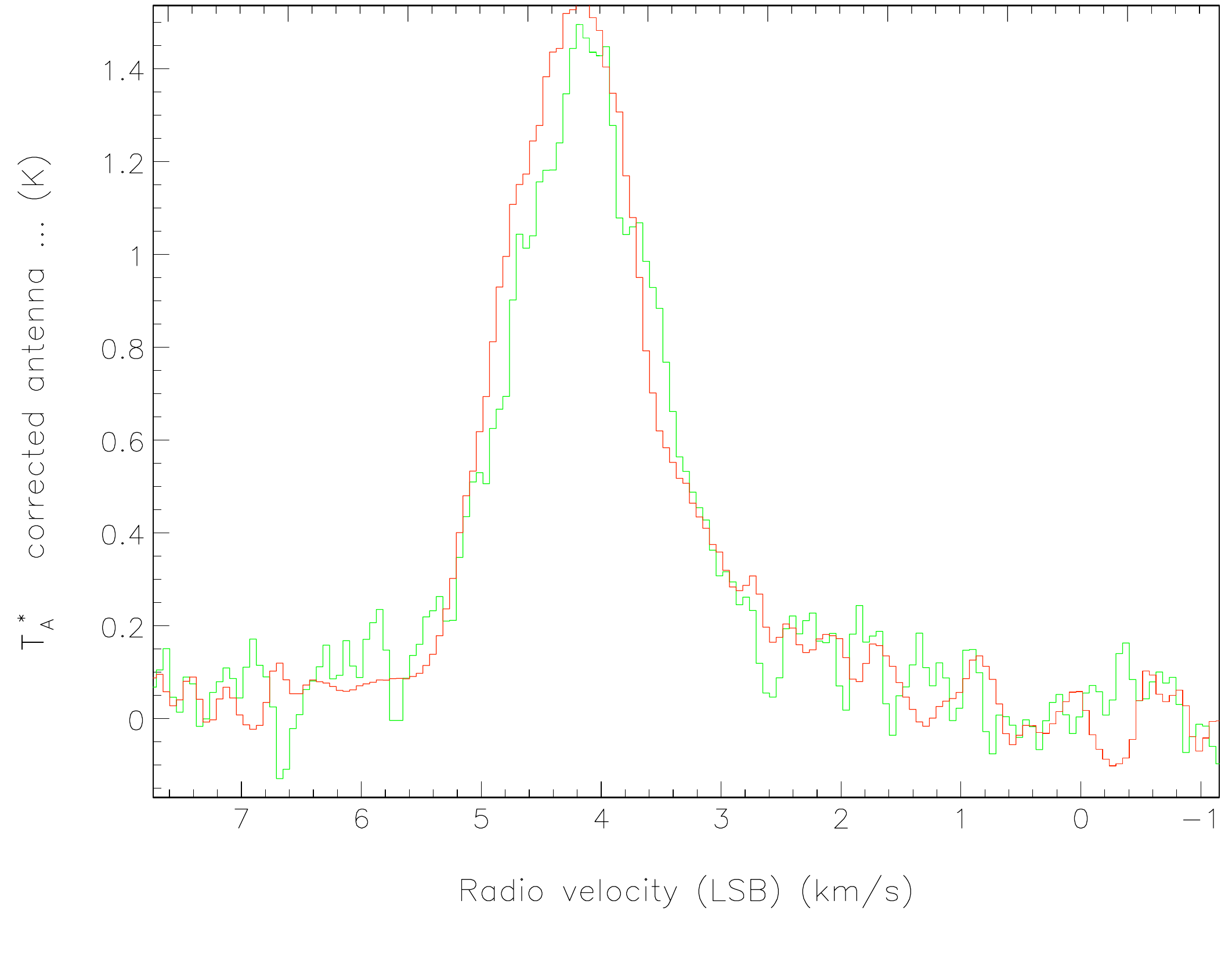}
\includegraphics[angle=-90,width=7cm]{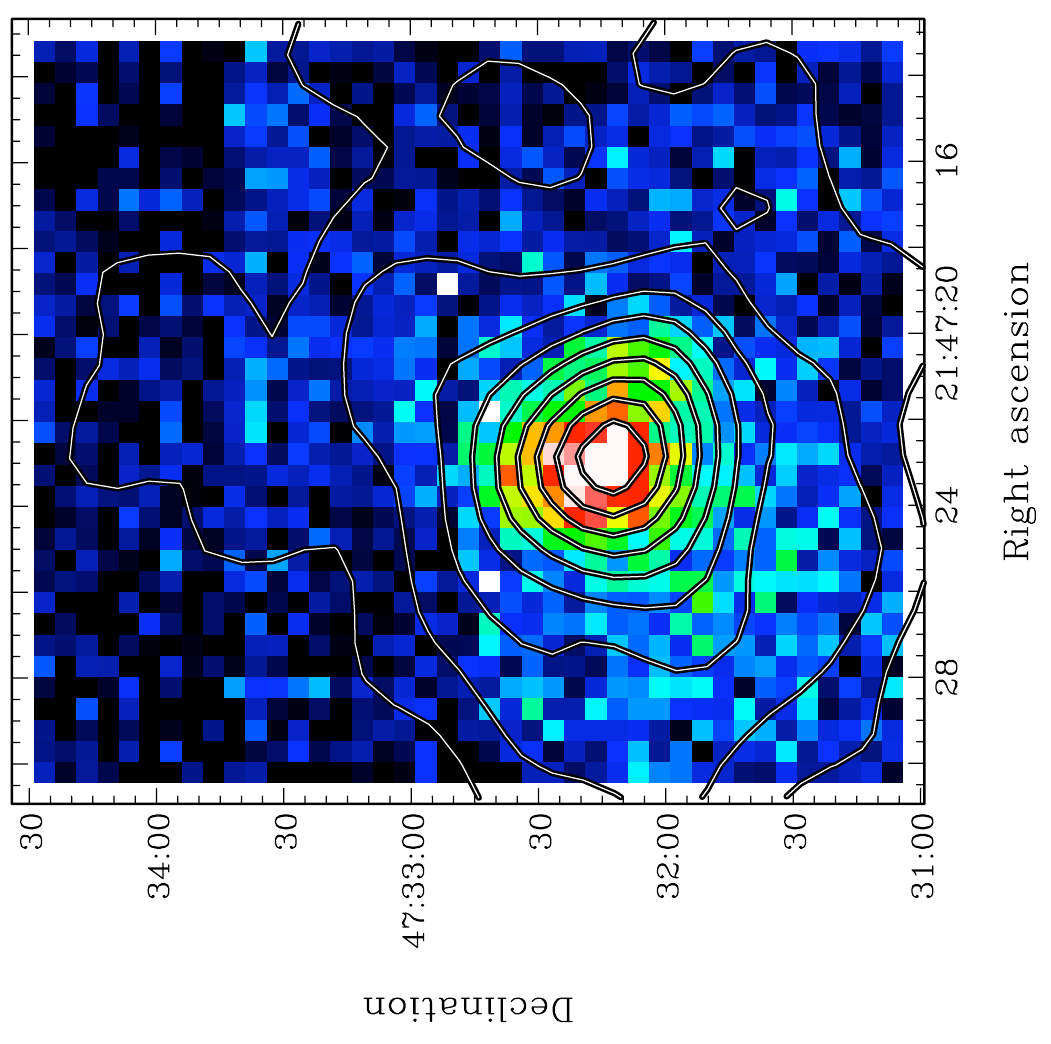}}
\caption{Left: Comparison of peak spectra for eastern core of IC\,5146. with RxB3 in red and HARP in green. Right: Comparison of the HARP data (contours)
with the previous receiver B3 observation (colour scale) of the eastern core of IC\,5146 in C$^{18}$O $J=$3 $\to$ 2.The contour levels levels are
of 0.25, 0.50, \ldots 2.25\,K\,km/s. \label{fig-ic5146}}
\end{figure*}

Finally, we have carried out a direct comparison of observations towards a protostellar core in IC\,5146, undertaken with HARP and with RxB3i. 
Fig.~\ref{fig-ic5146} shows the spectra of C$^{18}$O $J=$3$\to$2 taken with the two instruments at a single position, and the integrated intensity image, obtained with RxB3i, overlaid with contours from observations obtained with HARP. The RxB3i observations fully-sampled a region 180~$\times$200~arcsec in size, and took 10 hours of observing time. The HARP observations fully-sampled a region 1300~$\times$200~arcsec in size, and took 2.5 hours of observing time. The correspondence between the two sets of observations is consistently good both spatially and spectrally.

\section{Summary}
We have presented details of a new submillimetre spectral imaging system that has been installed and commissioned on the JCMT. HARP is a 16-element focal plane array receiver which operates from 325--375~GHz. Receiver temperatures are $\sim$120~K and under good weather conditions system temperatures are $\sim$300~K (SSB) at 345~GHz. 
At 345~GHz $\eta_{\rm mb}$ = 0.61, measured in 2007. A polarising Mach-Zehnder interferometer is used for sideband separation, and sideband rejection is better than 19~dB.
HARP operates in conjunction with ACSIS, which offers versatile bandwidth and spectral resolution configurations for all of the JCMT heterodyne receivers. In addition, ACSIS provides configurations that allow simultaneous observations of multiple lines, which can have different bandwidth and resolution configurations.

With HARP and ACSIS, the widest bandwidth available is 1.9~GHz with a channel spacing of 0.977~MHz, while the highest spectral resolution mode has a channel spacing of 31~kHz, or 0.03~\kms, with a bandwidth of 220~MHz. Since HARP is an array receiver, mapping speeds are increased, and it is possible to make fully-sampled maps of 1 square degree in less than 1 hour. A direct comparison between HARP and the previous JCMT single-pixel receiver RxB3i showed HARP observations mapping the region 28 times faster.

HARP and ACSIS provide simultaneous mapping and high resolution spectroscopy with high sensitivity, and with spatial dimensions well-matched to continuum images obtained with SCUBA. There are a wide range of scientific galactic and extra-galactic projects being carried out using this system at the JCMT. We presented some of the first science observations carried out as part of the HARP/ACSIS guaranteed time for the instrument teams, and highlighted the importance of observations using the new spectral imaging system in understanding astrophysical processes.

\section*{Acknowledgments}
The Cavendish Astrophysics Group would like to acknowledge the
technical expertize of P. Doherty, V. Quy and H. Stevenson in the
building of HARP. The authors would like to thank the referee, Gary Fuller, for helpful comments and suggestions. The James Clerk Maxwell Telescope is operated by The Joint Astronomy Centre on behalf of the Science and Technology Facilities Council of the United Kingdom, the Netherlands Organisation for Scientific Research, and the National Research Council of Canada.

\label{lastpage}

\end{document}